\documentclass[11pt,trackchanges]{aastex63}
\usepackage{natbib}
\usepackage{amssymb,amsbsy,amsmath}
\usepackage{verbatim}
\usepackage{times}
\usepackage{placeins}
\usepackage{aas_macros}
\usepackage[nottoc,numbib]{tocbibind}

\makeatletter
\def\set@curr@file#1{%
  \begingroup
    \escapechar\m@ne
    \xdef\@curr@file{\expandafter\string\csname #1\endcsname}%
  \endgroup
}
\def\quote@name#1{"\quote@@name#1\@gobble""}
\def\quote@@name#1"{#1\quote@@name}
\def\unquote@name#1{\quote@@name#1\@gobble"}
\makeatother

\usepackage{graphicx}

\newcommand{\sdo}{\textit{SDO}}

\received{\_\_\_\_\_\_\_\_\_\_}
\revised{\_\_\_\_\_\_\_\_\_\_}
\accepted{\_\_\_\_\_\_\_\_\_\_}

\submitjournal{ApJ}

\shorttitle{Fast, Robust Coronal DEMs}
\shortauthors{Plowman \& Caspi}

\begin{document}

\title{A Fast, Simple, Robust Algorithm for Coronal Temperature Reconstruction}
\correspondingauthor{Joseph Plowman}
\email{jplowman@nso.edu}

\author[0000-0001-7016-7226]{Joseph Plowman}
\affil{National Solar Observatory,
Boulder, CO 80303 USA}

\author[0000-0001-8702-8273]{Amir Caspi}
\affil{Southwest Research Institute,
Boulder, CO 80302 USA}


\begin{abstract}

We describe a new algorithm for reconstruction of Differential Emission Measures (DEMs) in the solar corona. Although a number of such algorithms currently exist, they can have difficulty converging for some cases, and can be complex, slow, or idiosyncratic in their output (i.e., their inversions can have features that are a result of the inversion code and instrument response, not of the solar source); we will document some of these issues in this paper. The new algorithm described here significantly reduces these drawbacks and is particularly notable for its simplicity; it is reproduced here, in full, on a single page. After we describe the algorithm, we compare its performance and fidelity with some prevalent methods. Although presented here for extreme ultraviolet (EUV) data, the algorithm is robust and extensible to any other wavelengths (e.g., X-rays) where the DEM treatment is valid.
\end{abstract}


\section{Introduction}\label{sec:intro}

The solar corona is filled with hot plasma, ranging in temperature from $\lesssim$1~MK to a few MK globally \citep[e.g.,][]{Schrijver1998, Sylwester2012}, up to $\sim$10~MK in dense loops above active regions \citep[e.g.,][]{Reale2009, Brosius2014, CaspiRocket2015}, and reaching up to $\sim$50~MK in intense solar flares \citep[e.g.,][]{caspisurvey2014, Warmuth2016}. While it is clear that the energy required to heat the plasma to these temperatures must come from the ubiquitous coronal magnetic field, the exact natures of the energy release and heating processes remain poorly understood \citep[e.g.,][and references therein]{Klimchuk2006, Holman2011}.

Different heating mechanisms yield different temperature distributions, with different spatial and temporal profiles. For example, in flares, ``direct'' heating of plasma in the corona is thought to yield high temperatures with rapid temporal variations, concentrated near the tops of the flaring loops \citep[e.g.,][]{Longcope2011, caspiimg2015, cheung2019}, while so-called ``chromospheric evaporation'' driven by impacting electron beams yields a cooler temperature distribution concentrated at loop footpoints and legs \citep[e.g.,][]{Holman2011, Allred2015}. In the quiescent corona, heating events are also likely impulsive \citep[e.g.,][]{Parker1988}, but ``flare-like'' magnetic reconnection-driven heating at low frequencies yields broad temperature distributions to high temperatures \citep[$\sim$10~MK, e.g.,][]{CargillKlimchuk2004, Cargill2014}, while high-frequency heating as might be expected from wave-driven processes yields narrower and significantly cooler temperature distributions \citep[e.g.,][]{AsgariTarghi2013}.  Accurately measuring the coronal temperature distribution is therefore crucial to distinguishing between and understanding these various heating mechanisms.

Hot coronal plasma emits across the entire electromagnetic spectrum, and this emission provides valuable diagnostics of the plasma temperature distribution \citep{Fletcher2011}. Collisionally-excited ions of various species emit narrow spectral lines at numerous wavelengths, from infrared through extreme ultraviolet (EUV) and soft X-rays, while free electrons interacting with these ions emit continuum emission from both (free-bound) radiative recombination and (free-free) bremsstrahlung processes, from radio to $\gamma$-rays. These emission processes and their dependence on temperature are well understood \citep[e.g.,][]{KochMotz1959, Landi2013} and measurements of the lines and continuum therefore serve to probe the coronal temperature distribution, or "differential emission measure" (DEM). Spectral lines in the EUV are especially bright compared to the photospheric background and have been particularly useful for imaging the corona in narrow passbands that include relatively isolated lines and hence sample fairly narrow bands in temperature \citep{Boerner2014}, such as implemented by the Atmospheric Imaging Assembly \citep[AIA;][]{Lemen2012} onboard the \textit{Solar Dynamics Observatory} \citep[\sdo;][]{Pesnell2012}. Consequently, a number of recent techniques have been developed to derive a spatially resolved coronal DEM from \sdo/AIA images in multiple passbands \citep[e.g.,][]{HannahKontar12, Plowmanetal_FIRDEM13, Cheungetal15, SuEtal_2018}.

Solving this DEM problem has become a frequently-tackled problem in solar physics, to the point where some might consider it a bit of an `old chestnut.' However, the existing algorithms are idiosyncratic in some respect or another, and tend to be complex. The \citet{Plowmanetal_FIRDEM13} algorithm, for instance, has difficulty recovering narrow input DEMs, and recovered broad DEMs tend to resemble the instrument temperature response function: this is a feature of the $L^2$ norm regularization (see Section \ref{sec:math}), and any algorithm relying on it is likely to have similar issues. The \citet{HannahKontar12} algorithm has difficulty effectively removing (unphysical) residual negative emission, and its solutions can contain excess high-temperature emission. The most promising recent algorithm, from \citet{Cheungetal15}, does well for a variety of cases, but requires tuning a somewhat arbitrary set of input basis elements and sometimes has difficulty with convergence of IDL's \texttt{simplex} algorithm, upon which it relies. 

This paper describes a new algorithm that mitigates all of these issues; it is particularly notable for its simplicity, and the IDL code is reproduced here on a single manuscript page. The algorithm guarantees positivity of the recovered DEM, enforces an explicit smoothness constraint, returns a featureless (flat) solution in the absence of information, and converges quickly to reduced $\chi^2$ of order unity -- and does so for all valid (i.e., consistent with the optically thin DEM assumption) inputs we have tested.

\section{Mathematics \& Description}\label{sec:math}

Spectral line emission from the solar corona is generally characterized as being optically thin, and for most lines can be written as

\begin{equation}\label{eq:losemission}
	I_{\lambda}(x,y) = \int \rho^2(x,y,z) G_{\lambda}(T(x,y,z)) dz\ \ \ \ \ \ \ \textrm{(photons)} ,
\end{equation}
where $x,y$ are in the plane of sky, $z$ is the line-of-sight direction, $\rho$ is the plasma number density, and $G_{\lambda}(T)$ is a plasma emissivity or `temperature response' function that is determined by the physics of the atomic transition line (and wavelength) in question. Cataloging a database of these functions is one of the purposes of the CHIANTI package \citep{Landi2013}. This can be generalized for a passband by adding together the temperature responses of each line contributing to the passband, weighted by the passband's transmissivity to the line, which we will write as $R(T)$, the result being equivalent to the $K(T)$ described in \cite{Boerner2014}. Because the corona is optically thin, these equations have no sensitivity to the arrangement of $\rho^2(x,y,z) R(T(x,y,z))$ along the line of sight, so, in practice $\rho(x,y,z)$ and $T(x,y,z)$ cannot be recovered from the observations. Instead, a related temperature function called a `differential emission measure' (DEM), ${\cal E}(T)$ can be inferred. It is defined in relation to Equation~\ref{eq:losemission} as

\begin{equation}\label{eq:dememission}
	I(x,y) \equiv \int {\cal E}(x,y,T) R(T) dT\ \ \ \ \ \ \ \textrm{(photons)} .
\end{equation}

Inverting this equation to find a coronal ${\cal E}(T)$ is an ill-posed problem since the temperature response functions are broad and may well be multi-peaked (e.g., for passbands containing multiple spectral lines that respond to widely separated temperatures). This means that additional constraints (such as smoothness) must be imposed for the solution to be mathematically well-defined (otherwise, singular matrices will generally result). The solutions must also be non-negative, as negative emission is non-physical, and absorption is an optically thick phenomenon which cannot be described by a DEM. 

\subsection{Linear Inverse Formalism}

To begin, we start with a set of observed data (typically in CCD data numbers) $D_i$, uncertainties $\sigma_i$, and temperature response functions $R_i(T)$ for a set of distinct spectral lines or passbands (for example, the \sdo/AIA 94, 131, 171, 193, 211, and 335 {\AA} channels, whose responses are available from \citealt{Boerner2014}) along a single line of sight (i.e., pixel location), whose emission is consistent with Equation \ref{eq:losemission}. For a given proposed DEM solution ${\cal E}(T)$, the `model' fits to the data, $M_i$, are given by Equation~\ref{eq:dememission} (dimensional differences between $I$ and $D$ are subsumed into the definition of $R_i(T)$):

\begin{equation}\label{eq:modeleqn}
	M_i \equiv \int {\cal E}(T) R_i(T) dT .
\end{equation}

We want to find the ${\cal E}(T)$ which produces the best fit to $D_i$ in the sense of minimizing the usual reduced $\chi^2$ statistic:
\begin{equation}\label{eq:chisquared}
	\chi^2 = \sum_i \frac{(D_i-M_i)^2}{\sigma^2_i} .
\end{equation}
To reduce the problem to a finite number of degrees of freedom and cast it in a form suitable for a linear algebraic treatment, we write ${\cal E}(T)$ as the sum of a finite number of basis functions, $B_j(T)$:

\begin{equation}
	{\cal E} = \sum_j c_j B_j(T) .
\end{equation}
The basis functions can, in principle, be any complete (down to some sampling limit) set, but the problem of enforcing positivity is most straightforward if they are each compact and localized at a distinct temperature. For example, \citet{warren2013} and \citet{caspidems2014} use Gaussian functions with fixed locations and widths in $\log_{10} T$.  Similarly, \cite{Plowmanetal_FIRDEM13} uses a set of triangle functions, equally spaced in $\log_{10} T$ with widths equal to twice their spacing; the implementation of our algorithm (see Appendix \ref{sec:implement}) will use the same basis. In any case, Equation~\ref{eq:modeleqn} then becomes

\begin{equation}
	M_i = \sum_j c_j \int R_i(T) B_j(T)dT \equiv \sum_j R_{ij}c_j ,
\end{equation}
thereby defining a `response matrix' $R_{ij}$ mapping the coefficients to the model fits to the data. Equation~\ref{eq:chisquared} then becomes

\begin{equation}\label{eq:chi2_respmatrix}
	\chi^2 = \sum_i \frac{\big(D_i-\sum_j R_{ij}c_j\big)^2}{\sigma^2_i} .
\end{equation}
The coefficients $c_j$ are thus the unknowns of the DEM solution we seek. The standard procedure here is to find the minimum of $\chi^2$ by solving for the zero of its gradient with respect to the $\bf{c}$ vector -- this being a quadratic form, the minimum is unique and the zero can be found analytically from

\begin{equation}\label{eq:chi2gradient}
	\frac{\partial\chi^2}{\partial c_k} = -2\sum_i\frac{D_i-\sum_j R_{ij}c_j}{\sigma^2_i}R_{ik} \equiv 0 .
\end{equation}
This equation is a form of the familiar linear inverse problem,

\begin{equation}
	{\bf b} = A\cdot {\bf c} .
\end{equation}
In this case, that standard form is achieved by defining

\begin{equation}
	{\cal R}_{ij} \equiv R_{ij}/\sigma_i ,
\end{equation}
\begin{equation}
	A_{jk} \equiv \sum_i{\cal R}_{ij} {\cal R}_{ik} , \quad\mathrm{and}
\end{equation}
\begin{equation}
	b_j \equiv \sum_i{\cal R}_{ij} D_i/\sigma_i .
\end{equation}
The coefficients are then given in vector form by

\begin{equation}\label{eq:sol_noreg}
	{\bf{c}} = A^{-1}\cdot{\bf{b}} ,
\end{equation}
At this point the ill-posed nature of the problem becomes evident. If the number of data channels (the first index of $R$) is fewer than the number of coefficients to be solved, $A$ is singular and $A^{-1}$ does not exist. 


One strategy to fix this, called `regularization,' is to add an additional constraint figure of merit to $\chi^2$. The simplest such constraint (mathematically) is to add a term proportional to the inner product of the coefficient vector, so that the figure of merit to minimize becomes

\begin{equation}\label{eq:l2normreg}
	\chi^2 \rightarrow F \equiv \chi^2+\lambda {\bf c}\cdot{\bf c} .
\end{equation}

This is sometimes known as Tikhonov regularization, or ridge regression, or the `$L^2$ norm', due to the constraint being proportional to $c^2$. Other constraints include the $L^1$ norm, which is proportional to $|c|^1$, and the $L^0$ norm, which is simply proportional to the number of coefficients. These lower order norms are more `sparse', meaning that they tend to result in solutions that span a smaller region of the solution space. It has been argued that this property tends to result in better solutions, in the absence of more specific constraints on the problem. However, these constraints can be weaker (e.g., in the case where two coefficients have the same effect on the data, the $L^1$ and $L^0$ norms will be singular, whereas the $L^2$ norm will not), and more difficult to solve for. See \cite{Cheungetal15} for references and more detailed discussion of these alternative norms. 


In the derivation of this method, we will continue from our definition of the $L^2$ norm (Equation \ref{eq:l2normreg}), although we will return to the question later in the paper (see Section \ref{sec:moreconstraints}). Solving for the minimum of $F$ is very similar to before, and the result is the same as Equation~\ref{eq:sol_noreg}, except that we make the substitution 

\begin{equation}\label{eq:tikhreg}
	A \rightarrow A+\lambda\bf{1} ,
\end{equation}
where $\bf{1}$ is the identity matrix (this makes $A$ manifestly nonsingular, since it is a positive matrix).  $\lambda$ is a parameter controlling the strength of the secondary constraint -- a larger $\lambda$ will result in a smaller $|{\bf c}|^2$ at the expense of a larger $\chi^2$, so $\lambda$ is often optimized such that the solution has $\chi^2\approx1$. Another, slightly more complicated constraint to minimize is the total of the squared derivative of the solution,

\begin{equation}\label{eq:derivconstraint}
	\gamma\int \Big[\frac{d {\cal E}}{dT}\Big]^2 dT = \gamma\sum_{ij}c_i c_j\int \frac{d B_i}{dT}\frac{d B_j}{dT} dT ,
\end{equation}
where, similar to $\lambda$, $\gamma$ is a tunable Lagrange multiplier controlling the strength of the regularization. It can be viewed as setting an upper threshold on the size of the derivative of the solution. Like before, the result is the same as Equation~\ref{eq:sol_noreg}, but this time the substitution is

\begin{equation}\label{eq:tikhreg_expand}
	A_{ij} \rightarrow A_{ij}+\gamma \int \frac{d B_i}{dT}\frac{d B_j}{dT} dT .
\end{equation}

It's pleasing to note that both of these solutions can be found in a single mathematical step: a simple linear matrix inverse (Equation~\ref{eq:sol_noreg}) is all that's required, provided optimal values of $\lambda$ or $\gamma$ can be found {\em a priori}. However, such solutions do not in general guarantee positivity, which is an essential component of a valid DEM solution. Even though the \cite{HannahKontar12} DEM method attempts to find a positive solution by simply increasing $\lambda$, such a procedure does not guarantee a non-negative solution (except in the trivial case of $\lambda = \infty$) and in the cases where such a solution exists it will often be over-regularized (i.e., $\chi^2$ significantly larger than unity). As a result, the solution will not match the data as well as it could and may miss important features of the source DEM which are discernable in the data. 

\subsection{Nonlinear Mapping and Inverse Formalism}

A more targeted means of guaranteeing positivity is called for; the \cite{Plowmanetal_FIRDEM13} technique uses an iteration for which this is arguably the case, but it fails to converge (or converges very slowly) for narrow DEMs. The DEM method described in this paper uses a more robust approach: to ensure positivity, we change to new variables, $s_{j}$, for the search, such that

\begin{equation}\label{eq:kappamapping}
	c_j = e^{s_{j}} .
\end{equation}
The DEM is then given by 

\begin{equation}
	{\cal E} = \sum_j e^{s_j} B_j(T) .
\end{equation}
This ensures that, provided the $s_j$ are real-valued and the basis functions are non-negative, the DEM solution will always be positive-definite. This exponential mapping works well for the DEM problem, but other choices which have the same positive-definite property, such as  $c_j = s_j^2$, may instead be made \citep[the specific choice, using the square, would lead naturally to an $L^1$ norm regularization, \'{a} la][]{Cheungetal15}. It's tempting to plug Equation~\ref{eq:kappamapping} directly into Equation~\ref{eq:chi2_respmatrix} for $\chi^2$  and find its minimum with the same procedure as before, but the resulting solution is identical to Equation~\ref{eq:sol_noreg} -- it simply returns imaginary-valued $s_j$ wherever $c_j$ is negative. To confine the $s_j$ to the real line, we Taylor-expand $c_j$ about some fiducial value, $s_j^0$:

\begin{equation}\label{eq:taylorexpansion}
	c_j \approx c_j(s_j^0)+(s_j-s_j^0)\left.\frac{d c_j}{d s_j}\right\vert_{s_j^0} \equiv c_j(s_j^0)+(s_j-s_j^0)c_j^\prime(s_j^0) .
\end{equation}
This can be plugged into Equation~\ref{eq:chi2_respmatrix} and $\chi^2$ minimized as before, keeping only first-order terms in $s_j$; Equation~\ref{eq:chi2gradient} becomes

\begin{equation}
	\frac{\partial \chi^2}{\partial s_k} = -2 \sum_i \frac{D_i - \sum_j R_{ij}\big[c_j(s_j^0)+(s_j-s_j^0)c_j^\prime(s_j^0)\big]}{\sigma_i^2}c_k^\prime(s_k^0) R_{ik} \equiv 0 .
\end{equation}
Grouping terms of the same order in the $s_j$ and solving for them, we find a solution of the same form as Equation~\ref{eq:sol_noreg}:

\begin{equation}\label{eq:sol_kappa}
	{\bf{s}} = A^{-1}\cdot{\bf{b}} ,
\end{equation}
except this time we have

\begin{equation}
	{\cal R}_{ij} = c_j^\prime(s_j^0) R_{ij}/\sigma_i ,
\end{equation}
and with a correction subtracted from the data, so that 

\begin{equation}
	D_i \rightarrow D_i-\sum_j R_{ij}\big[c_j(s_j^0)-s_j^0c_j^\prime(s_j^0)\big] .
\end{equation}
${\bf b}$ is otherwise unchanged. The first-order Taylor expansion makes this only approximate, but it provides an iterative scheme by which an accurate solution may be approached: start with an initial guess for $\bf{s}^0$, solve for $\bf{s}$ using these equations, update $\bf{s}_0$ with this new value of $\bf{s}$, and repeat. See Appendix \ref{sec:implement} for implementation details.

This solution will still tend to be ill-posed, however, since it includes no regularization other than the positivity criterion. We therefore add the derivative constraint previously introduced, but in the (natural) log space:

\begin{equation}\label{eq:nlregconstraint}
	\gamma\sum_{ij}s_is_j\int\frac{dB_i}{dT}\frac{dB_j}{dT}dT \approx \gamma\int\Big[\frac{d\ln{\cal E}}{dT}\Big]^2dT .
\end{equation}
This approximately minimizes the derivative of the $\ln$ of the DEM solution. In the case of the triangle-function bases, the solutions are piecewise linear interpolants in the (natural) log space, with the $s_j$ as control points. Expressing the regularization function in terms of the $s_j$ in this way keeps the problem simple and makes the regularization part of the figure of merit a quadratic form in the (natural) log space. It changes the $A$ matrix in exactly the same way as before (given in Equation~\ref{eq:tikhreg_expand}).


There are good reasons for minimizing the derivative in logarithmic rather than linear space -- specifically, the change in the logarithm of a quantity is equivalent (to first order) to the {\em relative} change in the quantity (compare with the well known $d\ln{f}/dx = f^{-1}df/dx$). This regularization therefore depends on the relative change in the DEM coefficients from one temperature to the next, not on the magnitude of the temperature derivative of the DEM. The latter will change depending on the amplitude of the input DEM function (requiring the strength of the regularization to be adjusted), whereas the former will not; no additional fine-tuning of the regularization (to achieve good $\chi^2$) is needed, since the same regularization strength applies regardless of the input DEM amplitude.

At this point, the choice of the regularization strength, $\gamma$, should be discussed. We want to penalize values of $d\ln{\cal E}/dT$ above some threshold value, $\delta_0$. 
If all of the $d\ln{\cal E}/dT$ are uniformly equal to $\delta_0$, on the other hand, the solution is nominal so the regularization figure of merit should be the same as a nominal $\chi^2$. Typically, this is $\chi^2=N_d$, the number of data points in the problem. We therefore choose $\gamma$ so that
\begin{equation}
	\gamma\int\Big[\frac{d\ln{\cal E}}{dT}\Big]^2dT\rightarrow\gamma\int\big[\delta_0\big]^2dT=N_d ,
\end{equation}
Therefore,
\begin{equation}\label{eq:drv_con}
	\gamma = \frac{N_d}{\delta_0^2\Delta T} ,
\end{equation}
where $\Delta T$ is the width of the temperature range being used (in the coronal DEM problem, $\log_{10}(T/K)$, with $K$ being Kelvin units, is typically used as the temperature variable and the useful temperature range for AIA is 5.5 to 7.5 Dex for a $\Delta T$ of 2.0).

\subsection{Summary of Derivation}

In summary, given input data $D_i$, errors $\sigma_i$, instrument response $R_i(T)$, and basis functions $B_i(T)$, the DEM inversion method is defined by the following equations. The DEM function itself is given by

\begin{equation}
	{\cal E}(T) = \sum_j c_j B_j(T) \equiv \sum_j e^{s_j} B_j(T) .
\end{equation}
Given an initial guess $s_j^0$, the coefficients $s_j$ are solved by inverting 

\begin{equation}\label{eq:linearproblem}
	{\bf{b}} = A\cdot{\bf{s}} ,
\end{equation}
with

\begin{equation}\label{eq:amatrix}
	A_{jk} \equiv \frac{N_d}{\delta_0^2\Delta T} \int \frac{d B_i}{dT}\frac{d B_j}{dT} dT+\sum_i{\cal R}_{ij} {\cal R}_{ik} ,
\end{equation}
\begin{equation}
	b_j \equiv \sum_i{\cal R}_{ij} \frac{D_i-\sum_j R_{ij}\big[c_j(s_j^0)-s_j^0c_j^\prime(s_j^0)\big]}{\sigma_i} ,
\end{equation}
\begin{equation}
	{\cal R}_{ij} \equiv c_j^\prime(s_j^0) R_{ij}/\sigma_i , \quad\mathrm{and}
\end{equation}
\begin{equation}\label{eq:rijfinal}
	R_{ij} \equiv \int R_i(T) B_j(T)dT .
\end{equation}

This can be iterated from some reasonable initial guess, $s_j^0$ (the choice of which does not appear to affect the result), until a reasonable $\chi^2$ is achieved; Appendix \ref{sec:implement} details our implementation of the method. It also includes a complete listing of our algorithm in IDL, which fits in less than one page, highlighting its simplicity.

The parameter $\delta_0$ defines the strength of the regularization. It can be roughly thought of as the number of $e$-foldings allowed in the DEM without incurring a regularization penalty; smaller values imply a {\em stronger} constraint, with stiffer, smoother solutions. With our implementation, we find good solutions (reduced $\chi^2$ of order unity) for any value between 4 and 16 (larger values still have good $\chi^2$, but are not smooth), which suggests that the constraint is `orthogonal' in some sense to the information supplied by the AIA passbands: accordingly, we posit that AIA passbands contain very little information about variation in the slope of the DEM (i.e., its derivative with respect to temperature) smaller than the characteristic width of the temperature response functions. Consequently, there is no need to search for an optimal value of the regularization strength for AIA applications.

\section{Test Cases, Comparison, \& Discussion}\label{sec:test}

The performance of the algorithm is now illustrated with several test cases. Detailed comparison of existing algorithms with a wide variety of test cases is beyond the scope of this paper and merits a paper in its own right \citep[cf.][]{Aschwanden2015}, but the test cases here include comparison to one of the most widely used contemporary methods, that of \cite{Cheungetal15}, as well as to \citet{Plowmanetal_FIRDEM13}. We have also evaluated the performance of the \cite{HannahKontar12} algorithm and found that it performs similarly (or worse) than the \cite{Plowmanetal_FIRDEM13} algorithm (see that paper for some examples), and hence omit a detailed comparison here. We use the standard and most up-to-date version of the \cite{Cheungetal15} algorithm, with its default settings (in particular, the adaptive tolerance setting is enabled and basis sigmas of 0.0, 0.1, and 0.2 were used); we did briefly investigate the modifications from \cite{SuEtal_2018}, but did not see an improvement for the cases we checked.

\subsection {Synthetic Test Cases}

\begin{figure}[!htbp]
	\begin{center}\includegraphics[width=\textwidth]{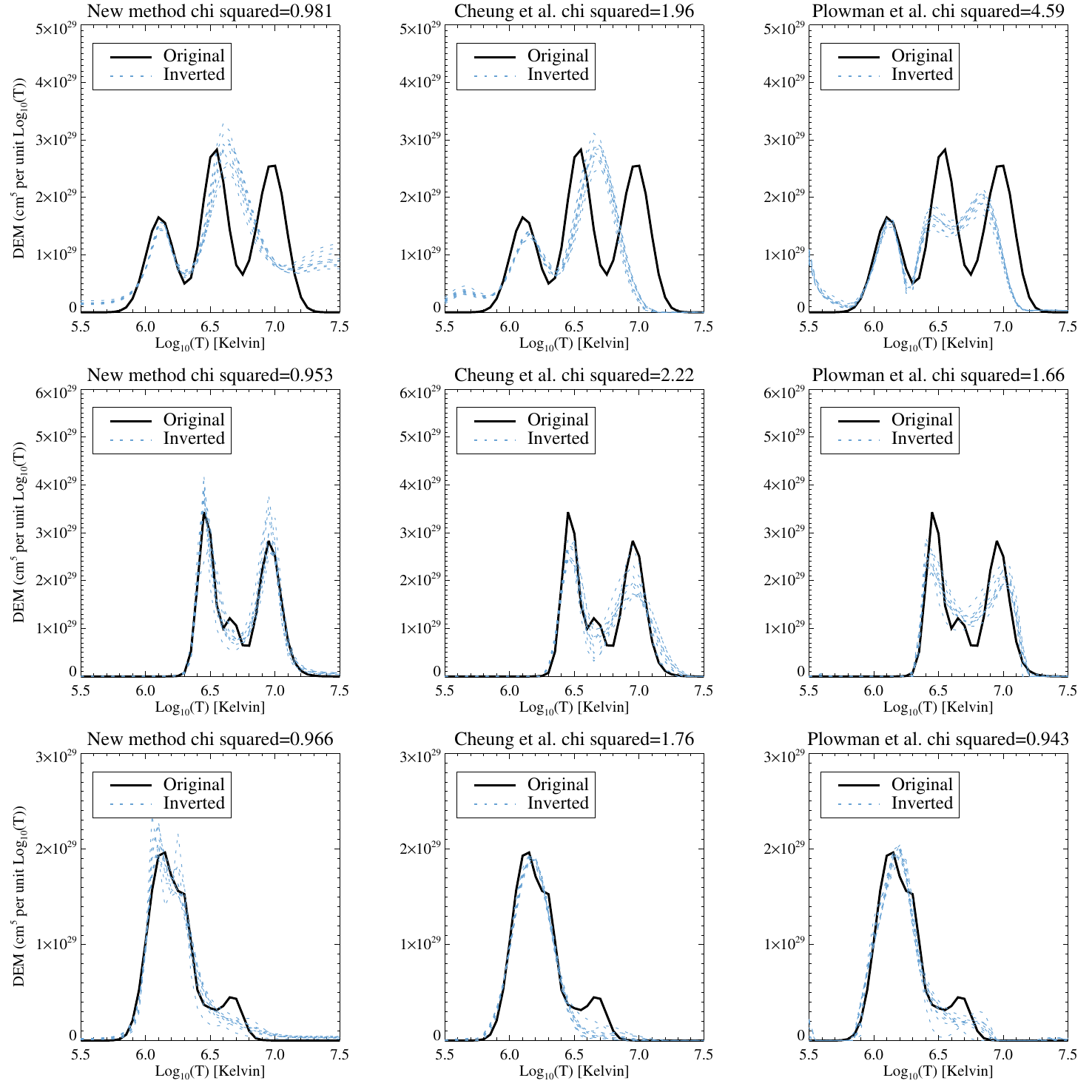}\end{center}
	\caption{Inversion of random temperature distributions. Each input DEM (solid) is the sum of 5 log-normal distributions with total emission measure between $5\times 10^{27}$ and $5\times 10^{28}~\mathrm{cm}^{-5}$ (uniformly distributed in $\log$ EM), standard deviations between 0.05 and 0.15, and central temperatures between $10^{6}$ and $10^{7}$~Kelvin.  The dashed lines show inversions with different random detection noise realizations. Each row compares the inversion of an identical input DEM between three methods: the new method described in this paper (left), the \citet{Cheungetal15} method (middle), and the \citet{Plowmanetal_FIRDEM13} method (right).}\label{fig:randomlognormals}
\end{figure}

Figure~\ref{fig:randomlognormals} compares the algorithms' inversion of three random DEMs. Each random DEM is a sum of 5 log-normal input DEMs, which have total emission measure between $5\times 10^{27}$ and $5\times 10^{28}$~cm$^{-5}$ (uniformly distributed in $\log$ EM), standard deviations between 0.05 and 0.15, and central temperatures between $10^{6.0}$ and $10^{7.0}$~Kelvin. In each case, the overall envelope of the distribution is well recovered, and generally if there are two well-separated peaks in the input, they are also separated in the recovered DEM. Smaller secondary peaks are generally not recoverable due to the temperature resolution and signal-to-noise of AIA, and triple-peaked distributions are recoverable only in ideal circumstances (not shown in these random test cases).  This is true of each algorithm.

It's important to note that the differences between the input and recovered DEMs are not due to failure of the algorithms to converge -- in each case, the reduced $\chi^2$ is of order unity, reflecting a reasonable fit to the data. They are also not due to the effects of instrument noise: each example in Figure~\ref{fig:randomlognormals} shows a set of reconstructions (dashed curves) of the same input DEM with different random instrument noise, to show the sensitivity of the reconstructed DEMs to such noise. In each case, there is some temperature range where none of the `bundle' of reconstructed DEM curves overlap the true solution; reducing the errors will simply shrink the thickness of the bundle about its mean, but will not bring it into agreement with the actual solution. These differences reflect the intrinsic ambiguity in the ill-posed inversion problem -- information is lost in the forward transform (Equation~\ref{eq:dememission}) and the preferences of the regularization must be substituted in its place as part of the reconstruction.

\begin{figure}[!htbp]
	\begin{center}\includegraphics[width=\textwidth]{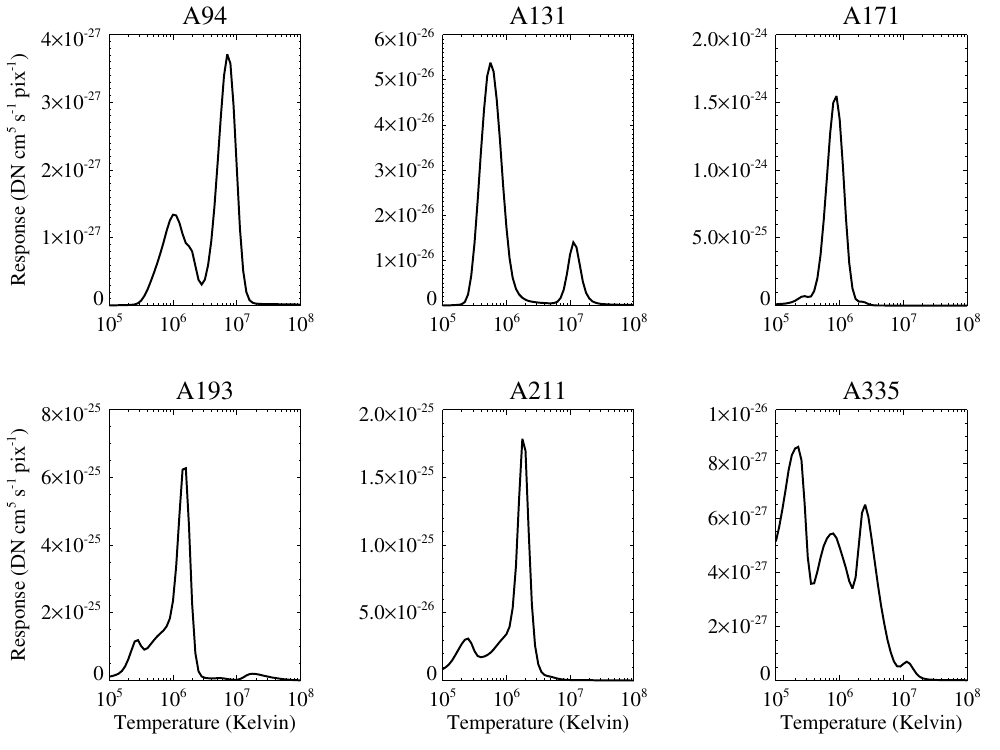}\end{center}
	\caption{Temperature response functions of the six AIA coronal channels (channel names are the wavelengths in Angstroms, prefixed with an `A').}\label{fig:aiatresp_multiplot}
\end{figure}

The most interesting of the test cases in Figure~\ref{fig:randomlognormals} to compare is the top row, and in particular its peak at 7 MK. We see that the \cite{Cheungetal15} method does not recover the peak at all, while the new method places additional emission measure there, but not in the form of a peak. This is because the new method prefers a flat DEM (in the absence of other constraints), while the \cite{Cheungetal15} method prefers to minimize the emission measure, and there is very little in the AIA temperature response functions (Figure~\ref{fig:aiatresp_multiplot}) to constrain the DEM above $10^{7.25}$~Kelvin. Morever, what little constraint is present is found in the AIA 193 \AA\ channel, which is also sensitive to emission at $\sim$1.5~MK. Emission above 10~MK is therefore less well constrained when there is also emission at $\sim$1.5~MK, as there is in this case. This is in contrast to the second test case, which is similar except that it does not have emission around 1.5~MK; there, both peaks are recovered by both our new algorithm and the \cite{Cheungetal15} algorithm, and there is no extra emission at 30~MK. In the absence of such a constraint, the new method will not rule out or suppress high temperature emission, but will instead prefer an emission measure that is constant in temperature (i.e., it prefers to extrapolate with a straight line).

In Section \ref{sec:moreconstraints}, we discuss some ways to suppress high-temperatures in the reconstruction if it is problematic for the specific application (e.g., calculation of energetics), but the simplest is simply to limit the inversion to the temperatures which are well constrained (i.e., $\sim$0.3 to $\sim$10~MK). This temperature range is roughly delineated by the locations of the highest and lowest temperature peaks in the AIA temperature response functions (Figure~\ref{fig:aiatresp_multiplot}). Limiting the maximum temperature of the inversion to 10--15~MK is safe as long as there is no emission above 10--15~MK, which is generally true except in the presence of flares. In that case the flare EM is large, and the data provide stronger constraints at high temperatures, so the temperature range of the inversion can be increased (however, we note that AIA tends to saturate during flares so that those DEMs cannot be reliably recovered, at least not on a per-pixel basis).

\begin{figure}[!htbp]
	\begin{center}\includegraphics[width=\textwidth]{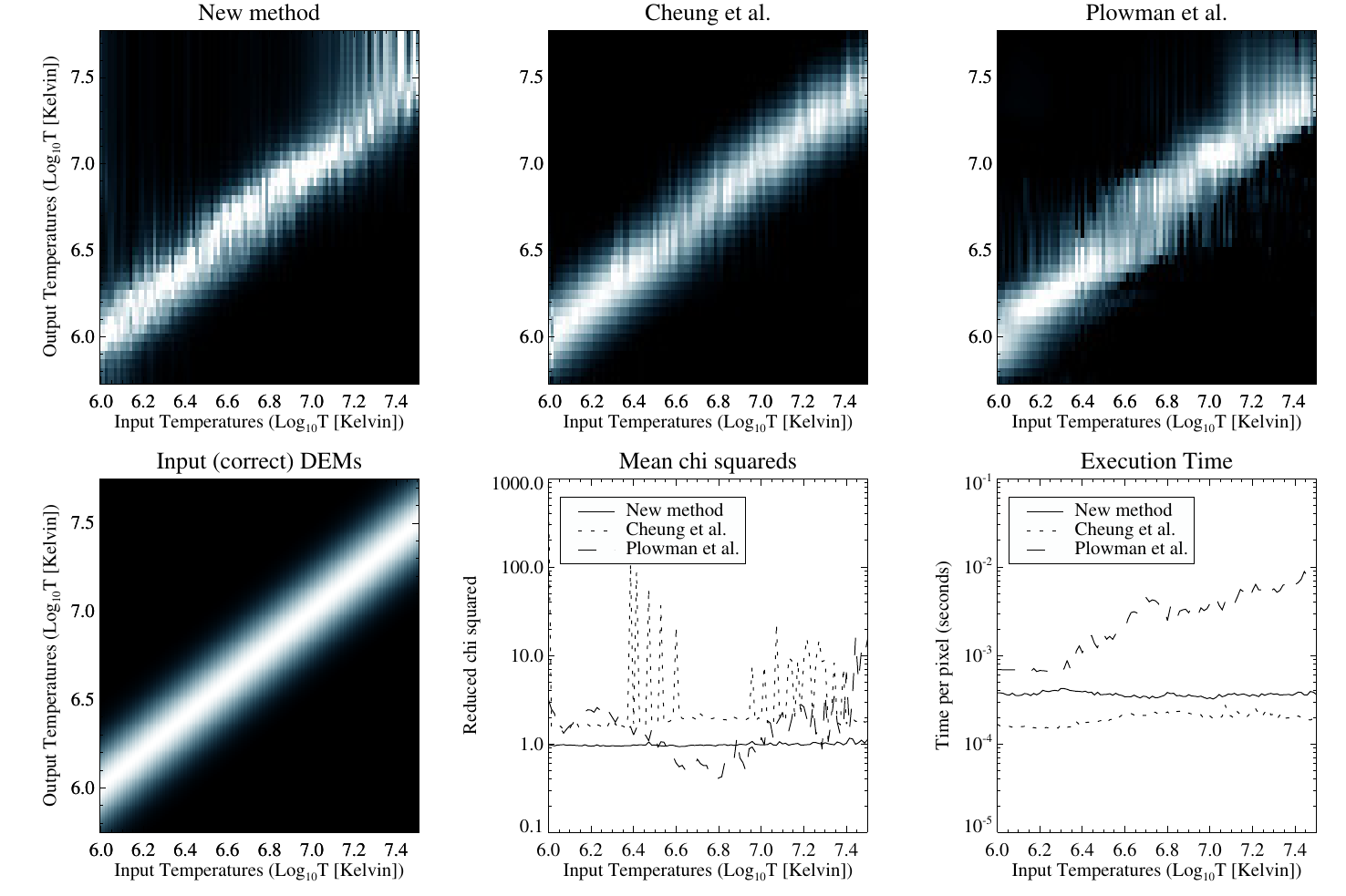}\end{center}
	\caption{Inversions of lognormal distributions with standard deviation of 0.15. These are concatenated into one image, with brightness showing the DEM intensity -- each vertical slice of the image is a DEM. The $x$ axis shows the temperature of each input DEM, which range from $10^{5.75}$ to $10^{7.25}$~Kelvin, while the $y$ axis shows the temperature variation of the DEM functions: Each vertical slice is a reconstruction of a log-normal input DEM centered at the temperature shown on the $x$ axis. The top row shows the inversions with three methods -- the new method described in this paper (left), the \citet{Cheungetal15} method (middle), and the \citet{Plowmanetal_FIRDEM13} method (right). The bottom row shows the input DEM functions which the inversions should ideally recover (on left), the reduced $\chi^2$ obtained by each algorithm (middle), and the execution time (right).}\label{fig:lognormal015}
\end{figure}

Next, Figure~\ref{fig:lognormal015} compares inversions of lognormal distributions with standard deviation of 0.15. In this case, the new algorithm recovers the shape of the input DEM, with variations due to noise in the input data and the ill-posed nature of the problem -- a smoother reconstruction can be obtained by changing the derivative constraint parameter (\texttt{drv\_con}) to a more restrictive value (from 8 to 1 or 2). This is reflected in the $\chi^2$ values all being $\lesssim$1. Comparison with the \cite{Cheungetal15} algorithm is a study in contrast -- although the reconstructed DEM is smoother and visually slightly closer to the ideal (see comment above), the average $\chi^2$ is considerably worse than our new algorithm, reflecting larger overall deviation. Additionally, although the \cite{Cheungetal15} algorithm is faster than our new code for this example case, that difference is smaller when the number of temperature points in the inversion is reduced; this inversion used 41 points from $10^{5.75}$ to $10^{7.75}$ Kelvin for testing and illustration, but we recommend 31 points with 10--15~MK as the maximum temperature for typical applications, only extending to $\sim$30~MK if necessary. 

The older \cite{Plowmanetal_FIRDEM13} algorithm fares somewhat worse: the core of the recovered distribution does not match the input as well, there is excess emission away from the peak, the recovered $\chi^2$ are poor in some cases, and execution times are longer (except near 1~MK). To avoid the impact of setup overhead on performance, each algorithm was tested on 100,000 data realizations with varying noise at each input DEM temperature.

\begin{figure}[!htbp]
	\begin{center}\includegraphics[width=\textwidth]{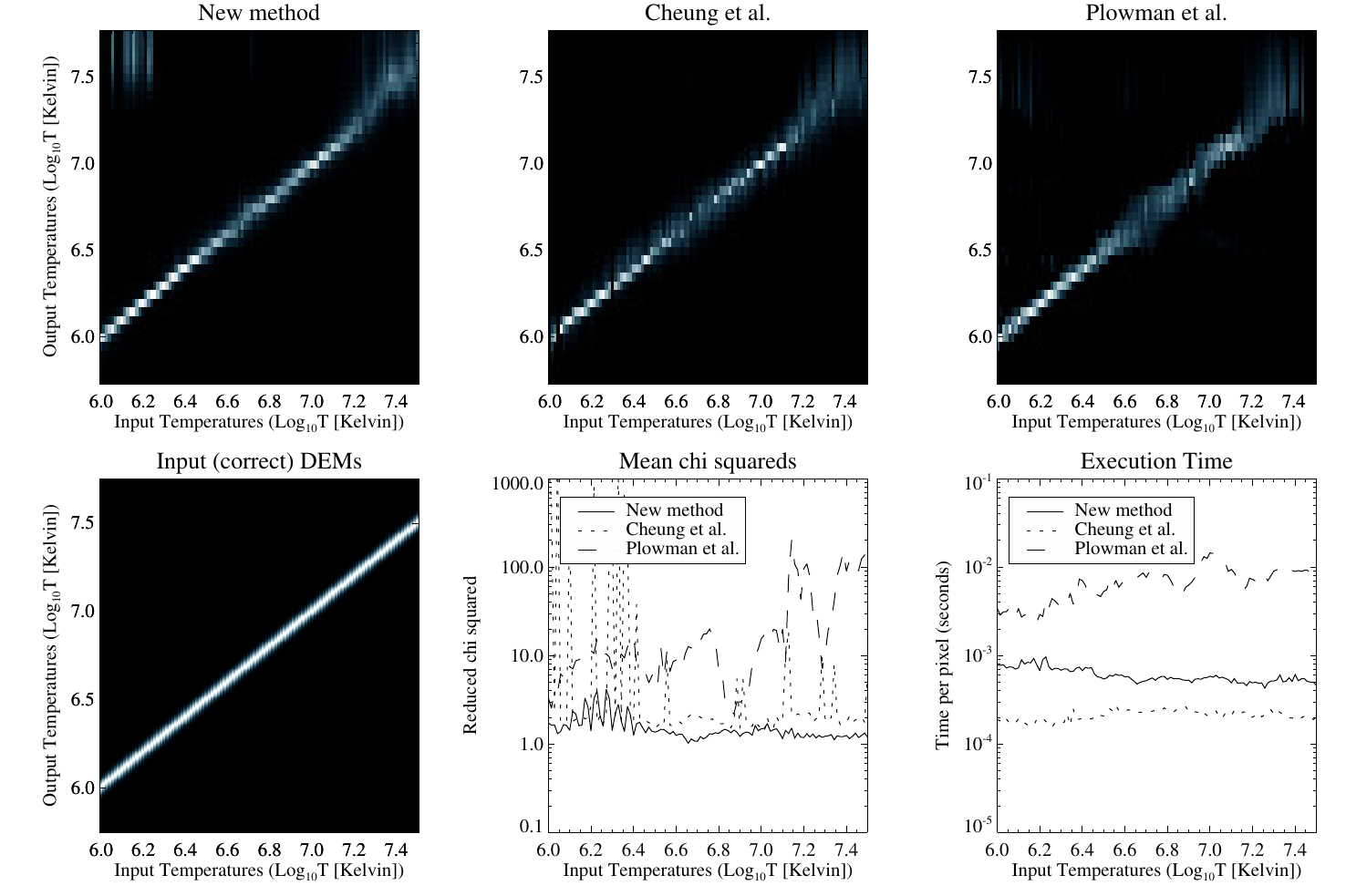}\end{center}
	\caption{Equivalent to Figure~\ref{fig:lognormal015}, but for lognormal standard deviations of 0.025 -- equal to the resolution of the AIA temperature response functions, and the basis functions of the DEM computation.}\label{fig:lognormal0025}
\end{figure}

Figure~\ref{fig:lognormal0025} is equivalent to Figure~\ref{fig:lognormal015} but compares inversions of nearly isothermal DEMs (lognormal with standard deviation 0.0025). These very sharp-featured DEMs are something of an `acid test' for this type of algorithm, which is designed to find solutions that are as smooth as possible. The widths of these DEMs are 0.05~dex, comparable to the basis function size (and the sampling resolution of the standard AIA temperature response functions), and yet for the most part the new algorithm produces reconstructed DEMs with minimal spurious features, fast performance, and good $\chi^2$.

The new algorithm does much better for this case than the \cite{Plowmanetal_FIRDEM13} method, which is affected by spurious features, has difficulty achieving good $\chi^2$, and has slower execution time. The \cite{Cheungetal15} method performs better than the \cite{Plowmanetal_FIRDEM13} method in this near isothermal case, but still significantly worse than the new method. The $\chi^2$ values are often poor and the method has difficulty finding solutions for some noise realizations (visible as gaps in the reconstructed DEM image, variation in $\chi^2$, and rapid alternation between narrow and broad solutions). 

The reconstructions for $\sim$1.5~MK input temperatures do contain some high-temperature ($\sim$30~MK) emission measure, accompanied by poorer (but still relatively good) $\chi^2$. This is not a failing of the DEM algorithm; rather, it is because the standard temperature resolution of our reconstructions (0.05~dex) is too low to reproduce the correct AIA passband emissions: when we recomputed the DEMs with higher temperature resolution (e.g., 0.025~dex, not shown), these spurious high temperature features went away and good $\chi^2$ was achieved. However, the ability to recover isothermal (or near-isothermal) DEMs only holds if they are in isolation, and it's debatable that our understanding of the temperature response functions is good enough to justify 0.025~dex temperature resolution; we feel that 0.05~dex respresents the best compromise between fidelity and performance.

Similar issues with excess high temperature emission can occur when the SNR is low; as in the multimodal example with emission at 1.5 MK (top row of Figure \ref{fig:randomlognormals}), ruling out $>$10 MK emission in the presence of 1.5 MK emission is difficult due to reliance on the AIA 193 \AA\ channel at high temperatures. For a similar reason, AIA has essentially no ability to distinguish between temperatures above $\sim$30~MK: only the 193~{\AA} channel is sensitive at those temperatures, and distinguishing between temperatures requires at least two channels. In these test inversions, we show temperatures up to 100~MK for reference, but any structure in the inversion at 100~MK is from the regularization and not from the data. In normal use (as already mentioned), we recommend limiting the upper end of the temperature range to 10--15~MK, unless there's good reason to expect the presence of high-temperature emission (e.g., a flare).



In each of these cases, the new algorithm converges quickly to a solution matching the input DEM (to within the limits of the ill-posed inversion) that has $\chi^2$ of order unity. The solutions are guaranteed to be positive, and the regularization constraint directly enforces smoothness while being largely independent of the data (i.e., the same regularization strength applies to a wide variety of DEM shapes and amplitudes). We now show inversions and comparison with real AIA data.


\subsection{Inversions with real AIA data}

\begin{figure}[!htbp]
	\begin{center}\includegraphics[width=0.5\textwidth]{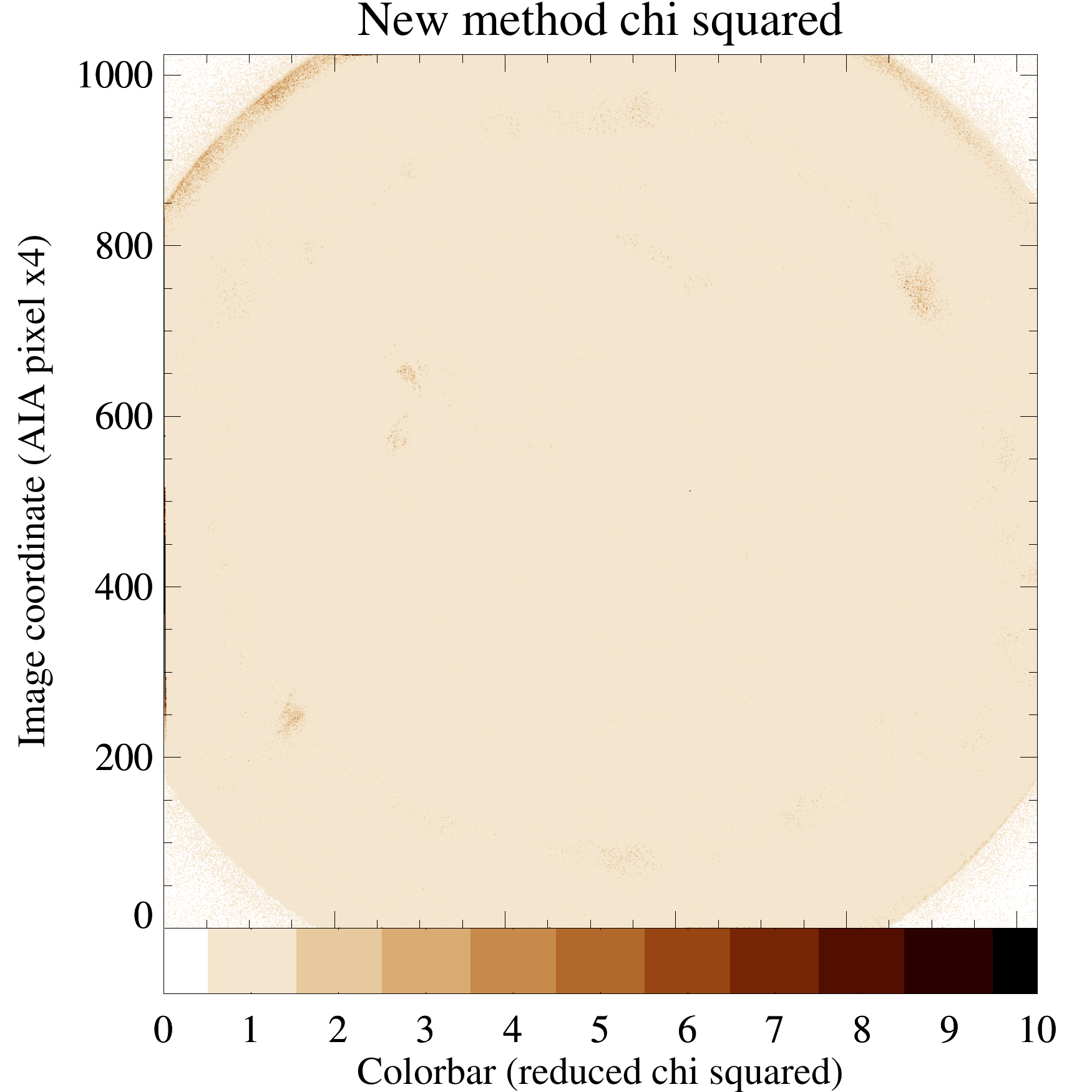}\includegraphics[width=0.5\textwidth]{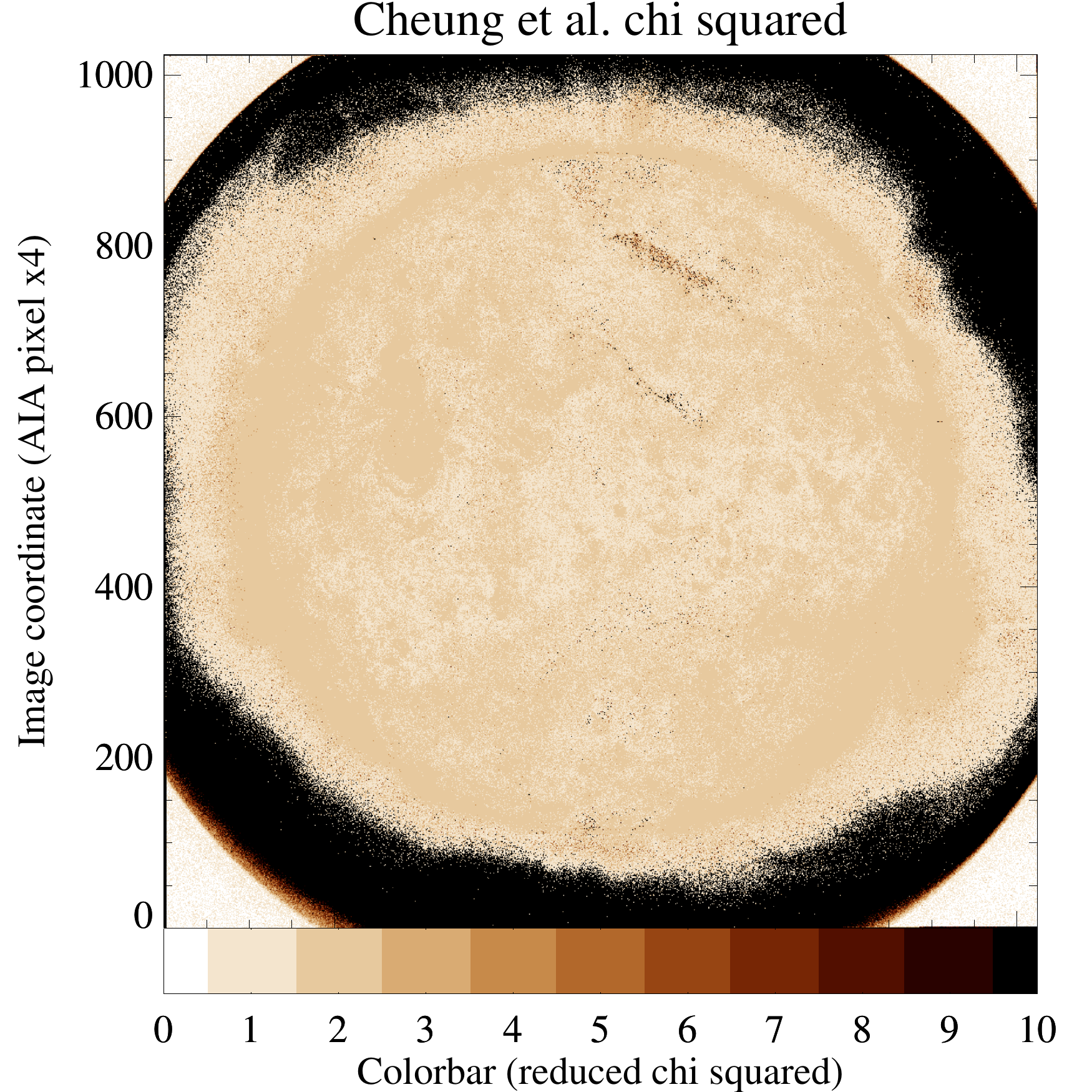}\end{center}
	\caption{$\chi^2$ for new method (left) compared with \cite{Cheungetal15} method (right), for the reconstructions in Figure~\ref{fig:dem_images}.}\label{fig:wholedisk_chi2}
\end{figure}

We have also compared our new algorithm with \cite{Cheungetal15} for a set of real {\sdo}/AIA data. The data were taken on 2010 August 1 at 06:30 UTC (the sequence shown for the coronal dimming, below, begins at this time, but it is before the initiation of flaring activity), and consist of one set of coronal frames (94, 131, 171, 193, 211, and 335 \AA ) spatially binned to 1024$\times$1024 pixels. We find that our method has good $\chi^2$ all the way out to the AIA field stop, and little residual structure is visible in the $\chi^2$ maps (Figure \ref{fig:wholedisk_chi2}). The \cite{Cheungetal15} method fares appreciably worse, with poorer $\chi^2$ in general, pixels where the algorithm failed to converge (especially off the disk where the noise is large). The \cite{Cheungetal15} method is also slower for this data than for the test cases, at $\sim$0.17 ms per pixel instead of $\sim$0.1 ms per pixel. Our new method, on the other hand, is faster for this data than for the test cases, at $\sim$0.2 ms per pixel (instead of 0.4), so the real-world performance of the two algorithms is roughly comparable.

\begin{figure}[!htbp]
	\begin{center}\includegraphics[width=0.5\textwidth]{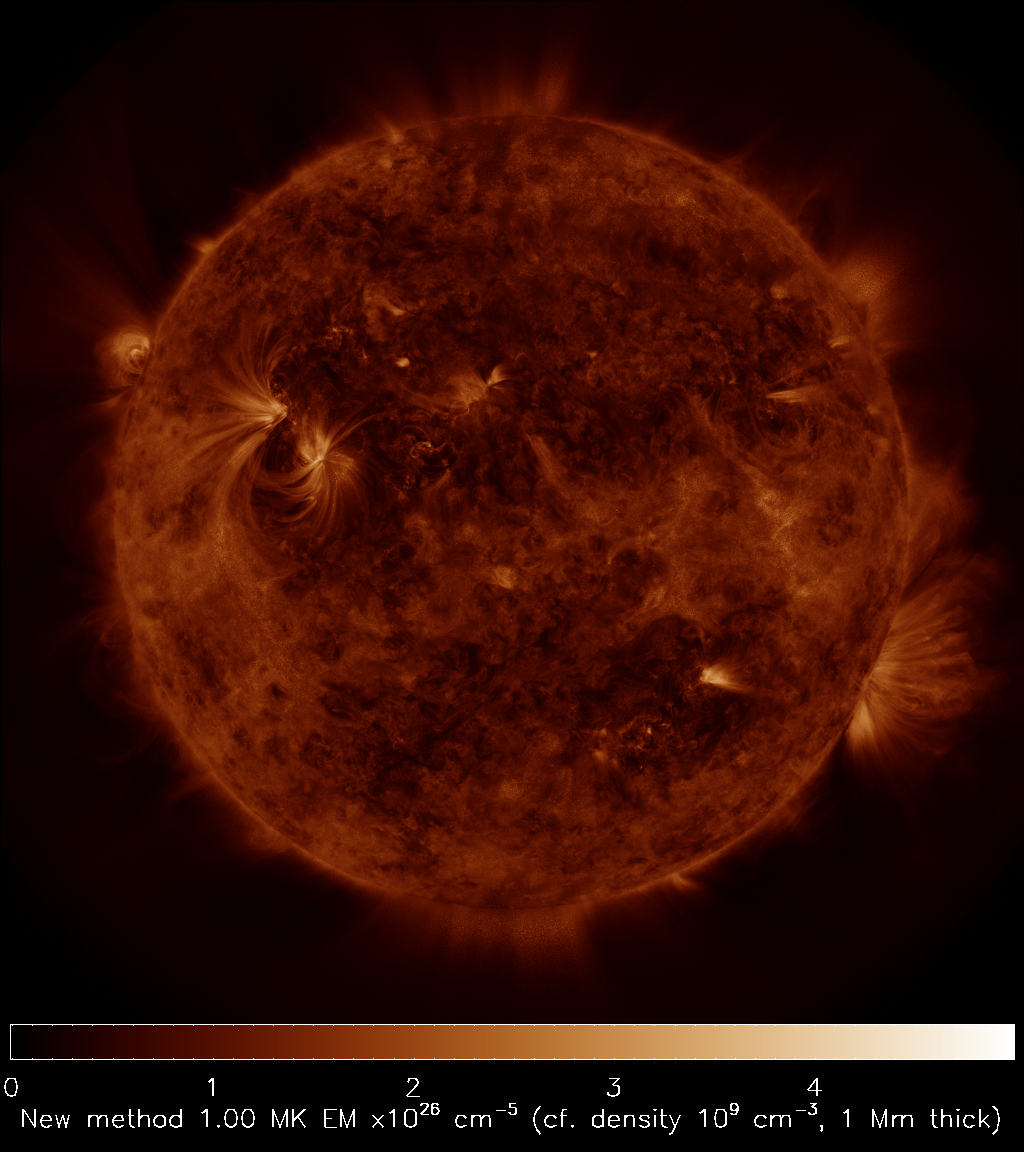}\includegraphics[width=0.5\textwidth]{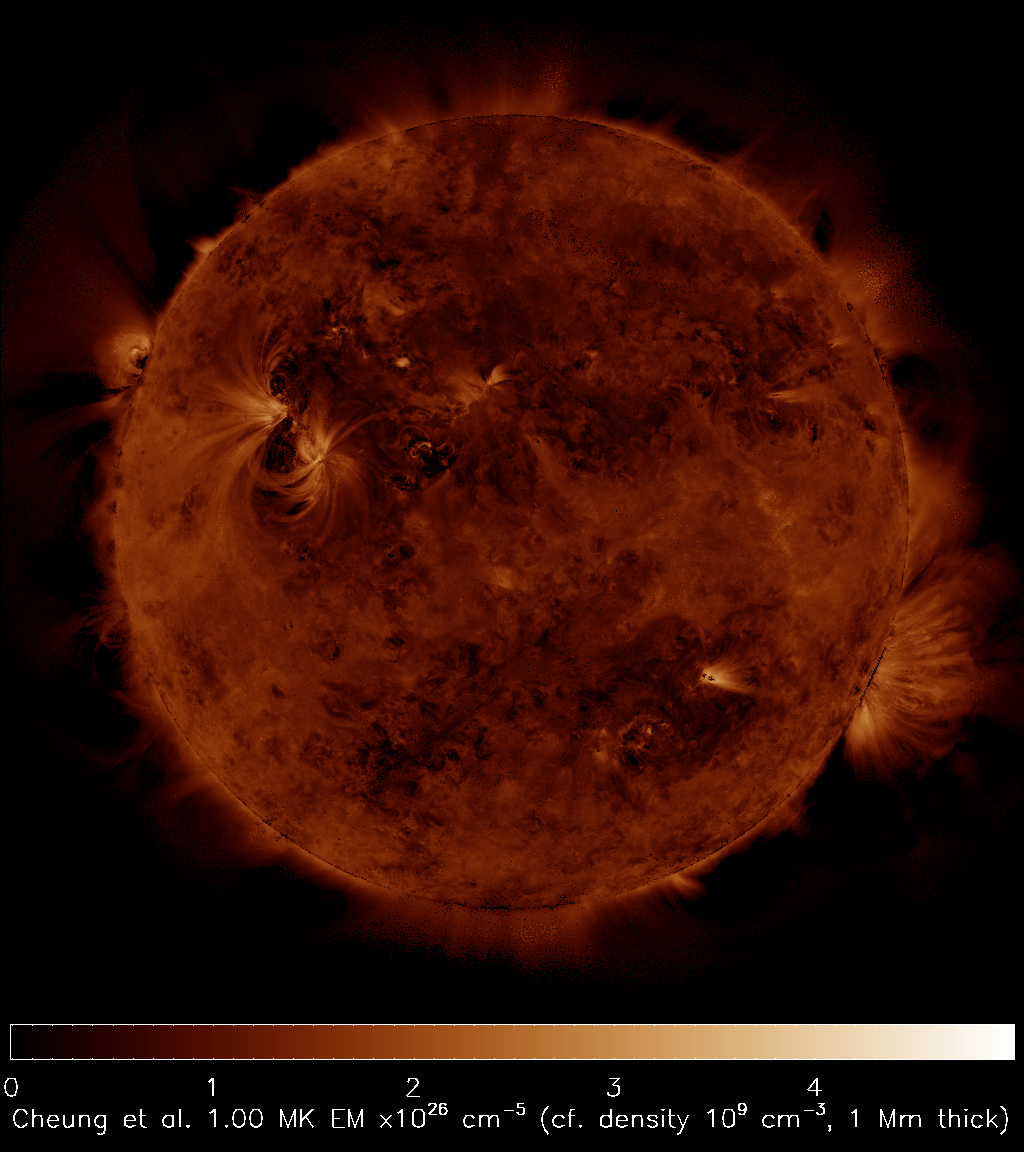}\end{center}
	\begin{center}\includegraphics[width=0.5\textwidth]{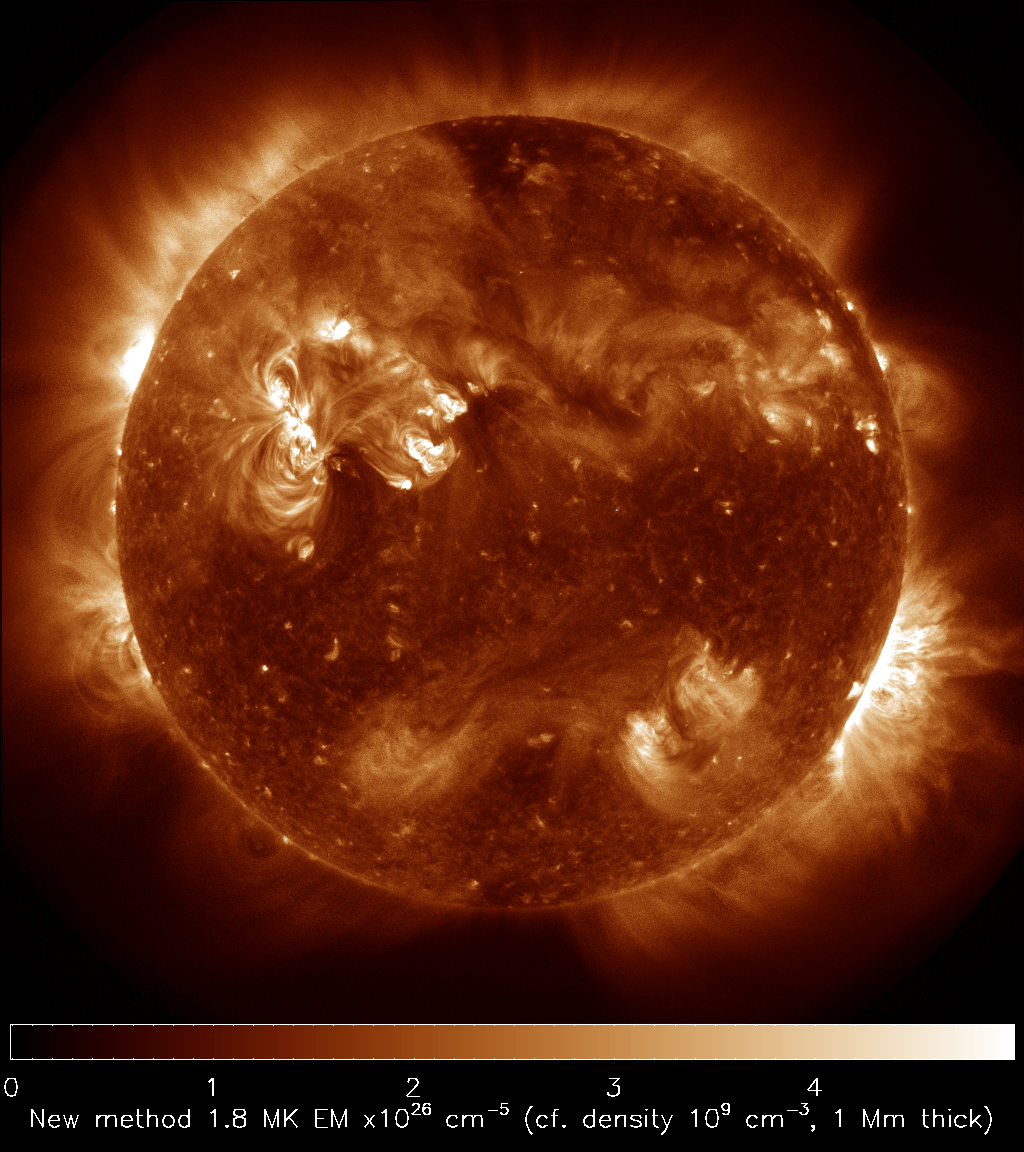}\includegraphics[width=0.5\textwidth]{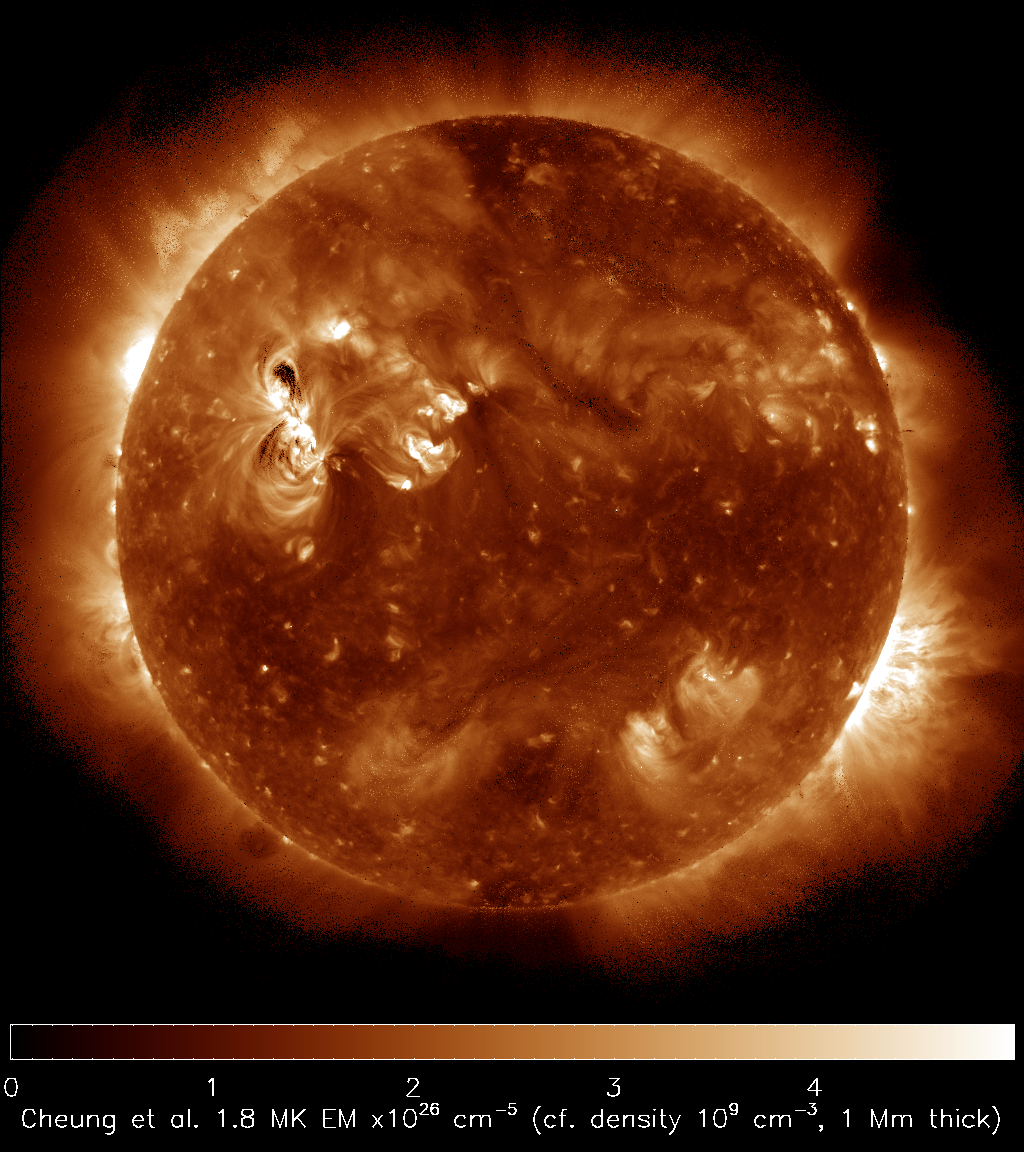}\end{center}
	\caption{Single-temperature slices of recovered DEMs from the new method (left) compared with the \cite{Cheungetal15} method (right), at 1 million Kelvin (top) and 1.8 million Kelvin (bottom). Minimum temperature of the inversion was 0.32 MK ($10^{5.5}$ Kelvin), maximum was 10 MK, and the derivative constraint (\texttt{drv\_con} in our code, $\gamma$ in Equation \ref{eq:drv_con}) was set to 8.}\label{fig:dem_images}
\end{figure}

The comparison is similar when we look at details of the reconstructed DEMs. The left panels of Figure \ref{fig:dem_images} shows the DEM computed from the new method at 1 million Kelvin and at 1.8 million Kelvin. Coherent structure, consistent the the measurements in the AIA passbands, can be seen all the way out to the field stop, and no artifacts from the DEM inversion are evident. The \cite{Cheungetal15} method (right panels of Figure \ref{fig:dem_images}), in contrast, shows artifacts at both temperatures -- pixel speckling and larger `holes' where the reconstruction shows no DEM at this temperature, and the faint structures at larger heights are not recovered. These differences are also reflected in the $\chi^2$ returned by the two algorithms (Figure \ref{fig:wholedisk_chi2}).



\subsection{Additional constraints, further examples}\label{sec:moreconstraints}

In certain cases, additional restrictions may be desired to constrain emission in regions where there is little instrument response (for example, we may want to include emission at $>$10~MK in the inversion for studying flares, but only where it is required by the data). For those cases, the data, errors, and temperature response functions can be supplemented with a constraint `channel'; the data is set to zero and the errors are set to a value which reflects a soft upper bound on the total emission measure. The algorithm will then try to minimize the emission in this channel, penalizing solutions above its `error' level. The constraint channel can be any desired function in general, but a simple and effective choice is a uniform function across temperature:

\begin{equation}
	R_\mathrm{const}=1.0
\end{equation}
This function gives the total emission measure when integrated against a DEM, and so, since our algorithm ensures DEM positivity, minimizes the total emission measure. It is therefore functionally equivalent to the `sparseness' constraint of \citet{Cheungetal15}. The soft upper bound is set to a conservative estimate for the total emission measure. Each of the channels or lines of a given pixel provides such an estimate, obtained by assuming that the emission comes from an isothermal DEM at its point of peak sensitivity (i.e., similar to traditional `EM loci' curves). The overall estimate for the total emission measure can then be taken to be the maximum of the individual estimates. Slightly more mathematically precise is to find the linear combination of the AIA channels which most closely approximates a constant temperature response function (i.e., by linear least squares), and use that linear combination as the estimate of the total emission measure. Either approach works reasonably well when the DEM falls between the peaks of the temperature response curves, but it can be too strict when it falls outside of them (e.g., for a flare), but we will use the linear combination method for the examples below. Estimating the total emission measure in this case is more difficult (akin to calculating the DEM in the first place), and we will refrain from following that rabbit trail here. 

\begin{figure}[!htbp]
	\begin{center}\includegraphics[width=\textwidth]{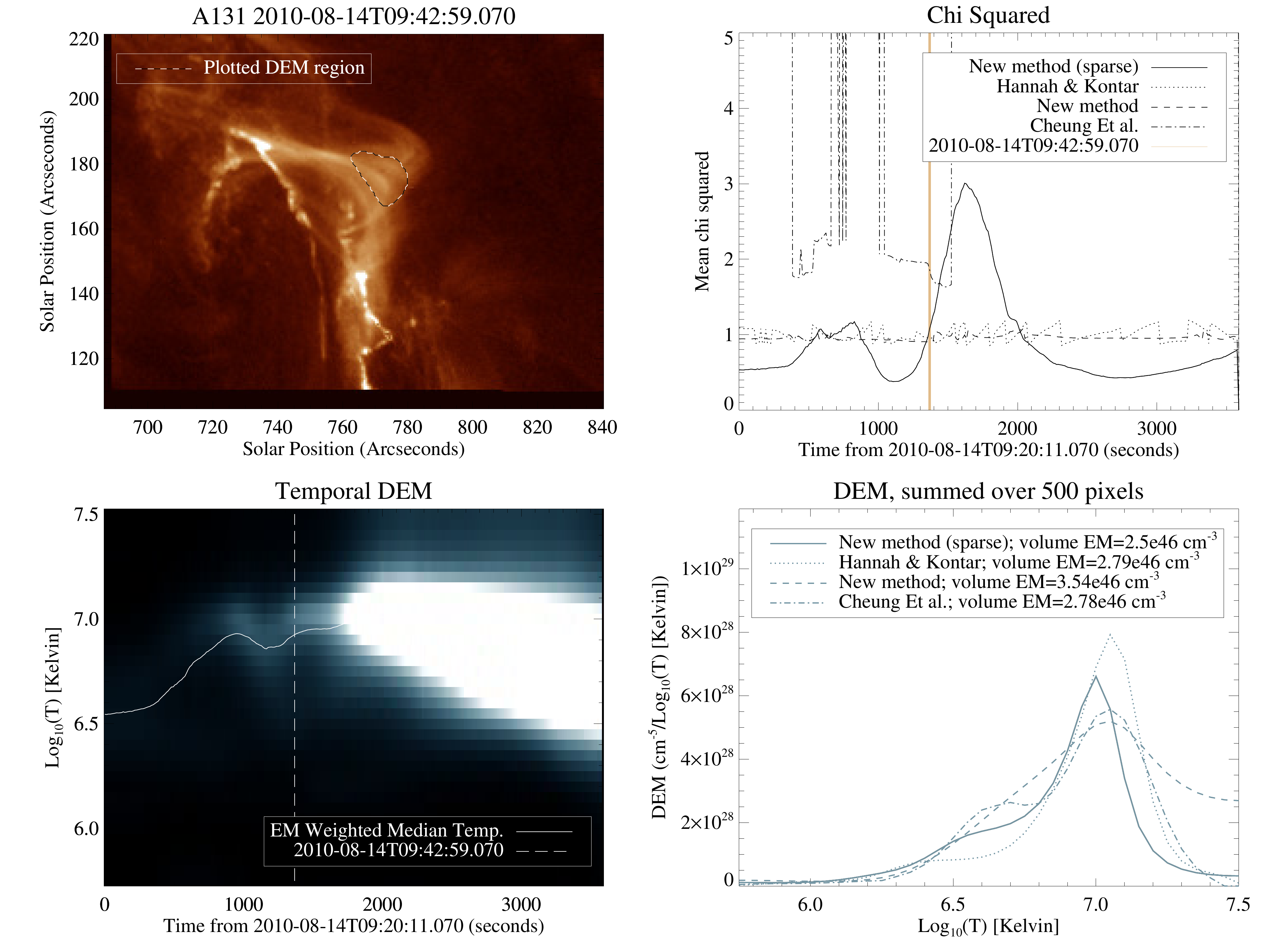}\end{center}
	\caption{Application of the DEM inversion method to the small flare studied by \citet{MotorinaKontar2015}. [Top left] AIA 131~{\AA} context image with contour indicating the region from which data were drawn to compute DEM. [Bottom left] Reconstructed DEMs using the new method, with sparsity constraint, computed from the region as a function of time (each vertical strip in intensity is a DEM). [Top right] $\chi^2$ achieved as a function of time. [Bottom right] Snapshot of the DEM at the time indicated by the dashed line in the $\chi^2$ and temporal DEM plots \citep[the same time as Figure~2 of][]{MotorinaKontar2015}.}\label{fig:mk2015data}
\end{figure}

To illustrate this constraint, and show the application of the algorithm to an existing data set in the literature, Figure~\ref{fig:mk2015data} shows inversion of solar flare data analyzed by \citet{MotorinaKontar2015}. The added constraint produces a somewhat lower DEM curve (especially at high temperature, as intended) at the cost of slightly higher (but still acceptable) $\chi^2$. A DEM curve computed using the \citet{HannahKontar12} method is also shown for comparison, which matches that shown in Figure~2 of \citet{MotorinaKontar2015} (this paper uses a different plotting convention; see Appendix~\ref{app:demconvention}). For this plot, we assume a systematic error of 20\%, to match \citet{MotorinaKontar2015}, in addition to added read and shot noise; the examples presented previously use read and shot noise only. This error level significantly limits the ability to recover details of the DEM (see discussion above), but is somewhat justifiable given the uncertainties in the atomic data and deviations of flare processes from the conditions assumed by DEMs (equilibrium, collisionally excited, no forbidden transitions, optically thin). The time series of the DEM shows that the emission measure observed by AIA peaks more than 10~minutes after the flare hard X-ray peak noted by \citet{MotorinaKontar2015} (at 09:46~UT); however, the high-temperature emission appears in the DEM around the time of the flare peak, and the peak temperature observed ($\sim$15~MK) also occurs at a similar time. This is similar to behavior observed in larger flares, where the highest-temperature emission (the so-called `super-hot' component) peaks at or near the flare hard X-ray peak, while cooler emission peaks later \citep[e.g.,][]{caspilin2010,caspisurvey2014,Warmuth2016}.

Other similar games can also be played with constraint `channels' -- e.g., the emission at high temperature can be `clamped' to be less than a particular value, ${\cal E}_\mathrm{clamp}$, by adding a `channel' which responds only to the highest temperature in the inversion (e.g., its temperature response is nonzero at the maximum temperature and zero elsewhere). Then, set the `data' for that channel to zero and its uncertainty to the data value the channel would record if the DEM at that temperature were ${\cal E}_\mathrm{clamp}$. 
The algorithm's derivative constraint will then attempt to smoothly match the rest of the DEM to this value. Of course, if there is actually high temperature emission being observed, this sort of constraint will erroneously suppress it in the recovered DEM.

One additional constraint worth mentioning is applying the $L^2$-norm style constraint, but in the log space -- i.e., of the form $\sum_i \ln(c_i)^2$. In the limit that the $c_i$ are large compared to unity, this approaches the $L^0$ norm (rather than the $L^2$ norm). In the DEM problem, the $c_i$ are quite large ($\sim$10$^{28}$) already, but in general the size of the $c_i$ can be controlled by scaling factors, allowing the effect (i.e., how much it is like the $L^0$ norm) to be tuned. An initial investigation of this idea suggests it has some promise, but will have to wait for a later work.

\begin{figure}[!htbp]
	\begin{center}\includegraphics[width=\textwidth]{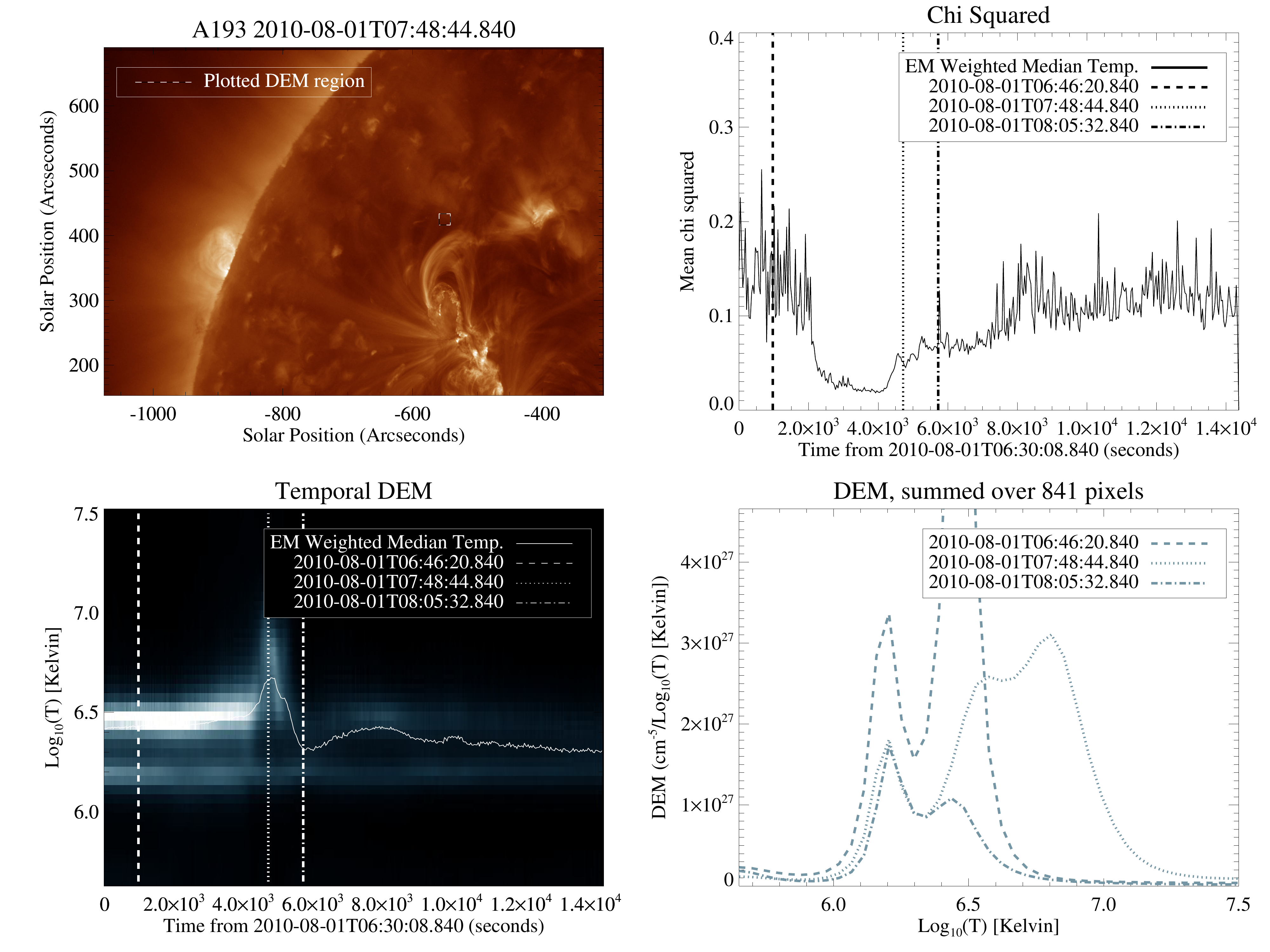}\end{center}
	\caption{New DEM method applied to a coronal dimming event. The format and arrangement of each panel is the same as for Figure~\ref{fig:mk2015data} but, in the bottom right snapshot, only the new method including sparseness constraint is shown.}\label{fig:dimmingregion}
\end{figure}

As a final example, Figure~\ref{fig:dimmingregion} shows the same type of analysis as Figure~\ref{fig:mk2015data}, but applied to a ``coronal dimming event'' \citep{masonetal2014,masonetal2016}. Here, the AIA emission in the coolest channels actually \textit{decreases}, in conjunction with a coronal mass ejection that removes a significant fraction of the emitting mass. This kind of analysis requires a DEM method that is fast and relatively free from idiosyncrasies, and is part of the motivation for the development of this algorithm. For cleanliness, only the inversions with the sparseness constraint are shown (this is important for this kind of event, since spurious high-temperature emission will obscure the heating and loss of material associated with dimmings). The region in question is indicated by the small dashed box in the upper left of the figure, and measures 30$\times$30 pixels; to increase signal-to-noise, the data are binned 3$\times$3 before the DEMs are computed, and these DEMs are then co-added at each time to produce the DEM time sequence shown on the lower left (this will typically produce a more detailed DEM than co-adding all of the pixels at the start). This DEM time sequence begins with a prominent emission feature at $10^{6.5}$~Kelvin, along with a secondary feature at $10^{6.2}$~Kelvin. The higher-temperature feature then appears to be heated to near $10^7$~Kelvin and subsequently vanishes as part of the dimming, while the lower-temperature feature persists but at lower emission measure: the implication is that the higher-temperature plasma -- possibly heated within the CME current sheet -- is `blown away' out of the active region during the CME, and much of the cooler mass is also removed by the CME, both of which result in the observed EUV dimming. Such EUV dimmings present an attractive means of characterizing CME mass as close to the acceleration region as possible \citep{masonetal2014}, but since EUV dimming can be caused by both mass loss and by temperature change (since each EUV passband has, in general, a narrow temperature response), an accurate reconstruction of the DEM in both time and space is crucial.  The positivity constraint from our new algorithm provides confidence in the DEM reconstruction across the temperature range, and also enables potential determination of CME mass through differences of pre- and post-eruption DEMs, which would not be well-defined with other reconstruction methods that do not enforce positive-definite DEM solutions at all temperatures.

\section{Conclusions}\label{sec:conclusions}

This paper describes a new method for inverting solar coronal differential emission measures. Although the examples shown are specific to AIA (perhaps the most common source of data for DEM inversions), it is straightforward to apply a variant of this algorithm to any data set for which the DEM assumption holds, such as for soft X-ray filter images from the \textit{Hinode} X-ray Telescope \citep[XRT;][]{Golub2008} or even for the Fourier-based imaging spectroscopy data from the \textit{Reuven Ramaty High Energy Solar Spectroscopic Imager} \citep[\textit{RHESSI};][]{Lin2002}. While intended for imaging data, the algorithm could easily be applied to spatially-integrated (``one-pixel'') spectral data, e.g., in X-rays from \textit{RHESSI} or from the \textit{Miniature X-ray Solar Spectrometer} \citep[\textit{MinXSS};][]{Woods2017, Moore2018} CubeSat, or in EUV from the {\sdo} EUV Variability Experiment \citep[EVE;][]{Woods2012} spectrometer.

This new algorithm has several features which recommend it for general use:

\begin{itemize}
	\item Its solutions are positive by construction, but despite the iteration required it converges quickly for a wide variety of input distributions.
	\item The log of the derivative of the DEM provides a constraint which is independent of its overall amplitude, making it more data-orthogonal than, for instance, regularization based on norms of the coefficient vector.
	\item It assumes no specific functional form for its solutions, allowing it to recover (for instance) multi-modal DEMs where data and regularization indicate it.
	\item A `sparsity'-based ($L^1$-norm) constraint is straightforward to add to the inversion if excess emission must be further minimized.
\end{itemize}

The algorithm's robustness and performance without requiring tuning to specific case make it well-suited to a variety of cases, especially those involving analysis of large volumes of solar data, and its performance in these respects is equal to or better than most available algorithms (a subsequent paper will compare the performance of a variety of algorithms in specific coronal applications). It is also very simple, and the robust convergence of the basic technique suggests it may be a good route forward to other ill-posed solar physics inverse problems.

The algorithm will be made available for general use in the SolarSoft \citep{Freeland98} IDL package as \texttt{simple\_reg\_dem.pro}, 
and a port for the python solar physics package, SunPy \citep{SunPyApJ,SunPyJOSS}, is in progress.

\acknowledgements{This work was funded by NASA grants NNX15AQ68G and 80NSSC17K0598. AC was also partially funded by NASA grants NNX14AH54G, NNX15AK26G, and 80NSSC19K0287. The authors thank Mark Cheung, Craig DeForest, Doug Nychka, and Dan Seaton, among others, for helpful discussions.}

\appendix

\section{Implementation}\label{sec:implement}
Our algorithm is implemented as a function in IDL. It takes as input an array of data images (`\texttt{data}'; $N_x\times N_y\times N_c$), corresponding arrays of uncertainty images (`\texttt{errors}'; $N_x\times N_y\times N_c$), and exposure times (`\texttt{exptimes}';  hereafter $\Delta t_i$, $N_c$ elements) for each channel or spectral line ($N_c$ channels in total). It also takes as input an ($N_T$ element one-dimensional) array of temperatures (`\texttt{logt}' for base 10 logarithm of temperature, although in principle it is insensitive to the choice of temperature variable), and an array of temperature response functions, `\texttt{tresp}' ($N_T\times N_c$) -- one for each channel or spectral line. For AIA, such arrays are contained in the structure returned by the SolarSoftWare (SSW) IDL routine `\texttt{aia\_get\_response}' as `\texttt{logte}' and `\texttt{all}', respectively, although care should be taken to omit the 304~{\AA} channel and truncate the temperature range to $\sim$5.5--7.5. It also accepts an argument (`\texttt{chi2}') which, on return, will contain an $N_x\times N_y$ array of $\chi^2$ values for each pixel (each argument/input are listed in order here). It returns the DEM solution (dimensions $N_x\times N_y\times N_T$). There are several optional keywords which control the behavior of the iteration and the regularization strength:
\begin{description}
	\item[\texttt{kmax}] The maximum number of iteration steps (default -- 100).
	\item[\texttt{kcon}] The number of initial steps to take before terminating due to $\chi^2$ failing to improve (default -- 15).
	\item[\texttt{steps}] Two-element array containing the large and small step sizes for the iteration (default -- 0.1 and 0.75).
	\item[\texttt{drv\_con}] The size of the derivative constraint -- threshold $\delta_0$ limiting the change in $\log$ DEM per unit input temperature (default -- 4; e.g., per unit $\log_{10}(T)$). See Equation~\ref{eq:drv_con} and accompanying discussion.
	\item[\texttt{chi2\_th}] Reduced $\chi^2$ threshold for termination (default -- 1).
	\item[\texttt{tol}] Terminate if $\chi^2$ improves by less than this from one step to the next (default -- $10^{-4}$).
\end{description}


\subsection{Setup}\label{sec:setup}

The initial lines of the code are concerned with checking input keywords, setting defaults, and finding the dimensions of the input arrays. After this bookkeeping, the next step is computing the matrix mapping the input DEM coefficients to the modeled output data, $R_{ij}$. This has a very close relationship to the array of input temperature response functions, which gives the value of the temperature response at a range of input temperature values. We will assume that the input temperature response functions are to be linearly interpolated between these values, that the basis functions have unit amplitude, are spaced according to the input temperatures, and that they are zero outside the input temperature range. Recall that the basis functions are assumed to be triangle functions, so they are given by:

\begin{equation}
B_j(T) = 
	\begin{cases}
		\frac{T-T_{j-1}}{T_j-T_{j-1}}, & T_{j-1}\leq T < T_j \ \&\ 0 < j < N_T \\ 
		\frac{T_{j+1}-T}{T_{j+1}-T_j}, & T_{j}\leq T < T_{j+1} \ \&\ 0 \leq j < N_T-1 
	\end{cases}
\end{equation}
That amounts to a linear interpolation scheme, and we interpret the input temperature response functions under the same scheme. This means that they can be expressed in terms of the same set of basis functions:

\begin{equation}
	R_{ij} = \int R_i(T)B_j(T)dT = \sum_k r_{ik}\int B_k(T)B_j(T)dT,
\end{equation}
where $r_{ik}\equiv  \Delta t_i R_i(T_k)$ is the transpose of the \texttt{tresp} array, scaled by the exposure times ($\Delta t_i$). The matrix enforcing the regularization (see Equations \ref{eq:nlregconstraint} and \ref{eq:amatrix}) is 

\begin{equation}
	{\cal D}_{jk}\equiv\int\frac{dB_j}{dT}\frac{dB_k}{dT}dT=\frac{\delta_{j-1, k-1}-\delta_{j, k+1}}{T_j-T_{j-1}}+\frac{\delta_{j+1, k+1} - \delta_{j, k-1}}{T_{j+1}-T_j}.
\end{equation}
These are straightforward to compute either analytically or numerically -- our algorithm uses the analytic route, which results in a faster and more compact setup. 

\subsection{DEM Computation}\label{sec:demcomp}

With the essential matrices computed, the code now loops over every pixel, computing the DEM for each. To begin the iteration, an initial guess is needed, and the code assumes a flat (uniform) DEM over the range of the input temperature response functions, ${\cal E}(T) = {\cal E}_0$. This avoids biasing the DEM toward particular temperatures and it begins the iteration at the minimum of the regularization. To choose the constant ${\cal E}_0$, the code uses a linear least squares fit. As previously mentioned, the DEM solution produced by the algorithm has little dependence on the initial guess.



The next step in the iteration can now be computed using Equations~\ref{eq:linearproblem} through \ref{eq:rijfinal}. For speedier execution of the time-intensive linear inverse step, a Cholesky decomposition is used rather than more general solvers; the matrix in question is symmetric by construction, so this is appropriate. The Cholesky decomposition will fail in the case of bad inputs (data, errors, or temperature response) -- pixels where this occurs will have $\chi^2$ set to $-1$ if the first iteration fails, and will have the value of the last successful iteration otherwise.

Stepping all the way to the new solution indicated by Equation~\ref{eq:linearproblem} leads to convergence problems because of overshooting in some cases, so the code only moves part of the way between the current solution and the new solution; after some trial and error, we found that a simple and effective solution is to try two different step sizes (10\% and 50\% of the distance between the two solutions, by default) and pick whichever one results in a lower $\chi^2$ as the new best solution. This generally leads to convergence in under $\sim$10~steps. Lastly, the code has convergence criteria for ending the iteration; this has three parts: 
\begin{enumerate}
	\item The algorithm will attempt to hit a target $\chi^2$ threshold, `\texttt{chi2\_th}' (default: 1), and will backtrack if it overshoots (i.e., $\chi^2$ becomes smaller than the threshold), to avoid overfitting. If the difference between the reduced $\chi^2$ and \texttt{chi2\_th} is less than a tolerance, `\texttt{tol}' (default: 0.1), then stop.
	\item If $\chi^2$ at the small step size fails to improve by more than a certain amount (`\texttt{tol}' times the small step size; default: 0.01), then stop. It was found that $\chi^2$ would sometimes increase during the first iterations (later converging back to $\sim$1), so this is only considered after some number of initial iterations (`\texttt{kcon}'; default: 5).
	\item If a maximum number of iterations (`\texttt{kmax}'; default: 100) have occured without $\chi^2$ convergence, then stop.
\end{enumerate}
The code proceeds to the next pixel once any of these are satisfied.

\newpage
\subsection{Code Listing}\label{sec:code}
\begin{footnotesize}
\begin{verbatim}
function simple_reg_dem, data, errors, exptimes, logt, tresps, chi2, $
        kmax=kmax, kcon=kcon, steps=steps, drv_con=drv_con, chi2_th=chi2_th, tol=tol

    if(n_elements(kmax) ne 1) then kmax = 100
    if(n_elements(kcon) ne 1) then kcon = 5
    if(n_elements(steps) lt 2) then steps = [0.1,0.5]
    if(n_elements(drv_con) ne 1) then drv_con = 8.0
    if(n_elements(chi2_th) ne 1) then chi2_th = 1.0
    if(n_elements(tol) ne 1) then tol = 0.1
    
    nt = n_elements(logt)
    nx = n_elements(data[*,0,0])
    ny = n_elements(data[0,*,0])
    dT = logt[1:nt-1]-logt[0:nt-2]
    Bij = (diag_matrix([0,dT]+[dT,0])*2.0 + shift(diag_matrix([0,dT]),-1) + $
        shift(diag_matrix([dT,0]),1))/6.0
    Rij = transpose(tresps*((1+dblarr(nt))#exptimes))#Bij ; Matrix mapping coefficients to data
    Dij = diag_matrix([0,1/dT]+[1/dT,0]) - shift(diag_matrix([0,1/dT]),-1) - $
        shift(diag_matrix([1/dT,0]),1)
    regmat = Dij*n_elements(exptimes)/(drv_con^2*(logt[nt-1]-logt[0]))
    rvec = total(Rij,2)

    dems=fltarr(nx,ny,nt)
    chi2=fltarr(nx,ny)-1.
    for i=0,nx-1 do begin
        for j=0,ny-1 do begin
            err = reform(errors[i,j,*])
            dat0 = reform(data[i,j,*]) > 0.0
            s = alog(total((rvec)*((dat0 > 1.0e-2)/err^2))/total((rvec/err)^2)/(1+dblarr(nt)))
            for k=0,kmax-1 do begin
                dat = (dat0-Rij#((1-s)*exp(s)))/err ; Correct data by f(s)-s*f'(s)...
                mmat = Rij*((1.0/err)#exp(s)) ; Weight mapping by 1/err and f'(s)...
                amat = transpose(mmat)#mmat+regmat
                la_choldc,amat,status=stat
                if(stat eq 0) then begin
                    c2p = mean((dat0-Rij#(exp(s)))^2.0/err^2)
                    deltas = la_cholsol(amat,transpose(mmat)#dat)-s
                    deltas *= (max(abs(deltas)) < 0.5/steps[0])/max(abs(deltas))
                    ds = 1-2*(c2p lt chi2_th) ; Direction sign; is chi^2 too large or too small?
                    c20 = mean((dat0-Rij#(exp(s+deltas*ds*steps[0])))^2.0/err^2)
                    c21 = mean((dat0-Rij#(exp(s+deltas*ds*steps[1])))^2.0/err^2)
                    interp_step = ((steps[0]*(c21-chi2_th)+steps[1]*(chi2_th-c20))/(c21-c20))
                    s += deltas*ds*((interp_step > steps[0]) < steps[1])
                    chi2[i,j] = mean((dat0-Rij#(exp(s)))^2.0/err^2)
                endif else break
                if((ds*(c2p-c20)/steps[0] lt tol)*(k gt kcon) or abs(chi2[i,j]-chi2_th) lt tol) then break
            endfor
            dems[i,j,*] = exp(s)
        endfor
    endfor
    return,dems 

end
\end{verbatim}
\end{footnotesize}

\section{On DEM Representation, Units, and Plotting Conventions}\label{app:demconvention}

Implicit in the expression for the emission in Equation~\ref{eq:losemission} is an integral over the weighted plane-of-sky area sampled by the point spread function (PSF) of the detector in question (e.g., a pixel in AIA), so that the DEM in Equation~\ref{eq:dememission} is differential not only in the emission along the line of sight but also within the plane-of-sky area; the two together form a volume from which the DEM is sampled. There are two conventions for dealing with this:

\begin{description}
	\item[Column emission measure convention] Normalize the DEM by the area covered (i.e., at the Sun's surface). If the emission originates from a uniform density patch of known column depth, the density can be found from the total (i.e., integrated) emission measure by dividing by the column depth and taking the square root (EM $= l\rho^2$). In this case, the pixel area is incorporated into the instrument response functions (i.e., the temperature response functions returned by the SSWIDL \texttt{aia\_get\_response(/temp,/dn)} have units of DN~cm$^3$~s$^{-1}$~cm$^2$/pixel).
	\item[Volume emission measure convention] Don't normalize by the area covered. In this case, if the emission originates from a uniform density patch of known column depth, the density can be found from the total emission measure by dividing by the volume and taking the square root (EM $= V/\rho^2$).
\end{description}

Each of these conventions sees use in the literature, and the conversion factor between them is the area of the region in question. In either case, it is the integral of the DEM over some temperature range that relates to the physical quantity -- the density: The integral of the DEM over some temperature range is the integral, over the volume in question, of $\rho^2$ at all points in the volume with temperatures in that range. dividing this integral by the size of the volume and taking the square root gives the root mean square (RMS) density in that temperature range, weighted by the instrument PSF.

Graphs and plotting of DEMs must therefore be careful not to distort this integral, which, if the DEM is properly plotted, is equivalent to the area under the DEM curve. The most important consideration in this case is that the temperature units of the DEM must match the temperature units of the $x$-axis (i.e., the temperature) of the plot: If the $x$-axis is logarithmic, then the DEM should be plotted per unit log of $T$ (typically $\log_{10}T$, with temperature in Kelvin). If the DEM per unit Kelvin is to be plotted, on the other hand, the temperature axis should be {\em linear}. Plotting the DEM per unit Kelvin with a logarithmic temperature axis leads to exaggeration on the low-temperature end of the temperature range -- the missing weighting is the conversion factor from $dT$ to $d\log_{10}{T}$, i.e., $T\log{10}$, so a feature at 0.5~MK will appear 20 times as large as one at 10~MK when each has the same total emission measure (and therefore the same density, all else being equal). In principle, either choice of temperature axis is appropriate as long as the DEM axis is consistent, but we would argue that a logarithmic temperature axis is more appropriate in practice: the spectral line emissivity curves from which temperature response functions are computed tend to have similar {\em relative} widths (a 10~MK line has $\sim$10 times the width of a 1~MK curve), and the recoverable resolution of DEMs scale with these widths. When plotted on a linear temperature axis, DEM features therefore tend to look narrow and tall on the low temperature end, broad and low on the high temperature end (assuming comparable total emission measure); the logarithmic temperature axis (and DEM plotted per unit $\log_{10}T$) normalizes such differences, and we use this plotting convention.

A less serious and more controversial choice is the $y$-axis of the DEM plot -- whether it should be linear or logarithmic. The convention in the field is to use a logarithmic scale on the $y$-axis, but the area argument made above favors a linear scale. Specifically, a log scale's compression of vertical range can make two curves whose integrals are very different have similar areas, and vice versa. And, although practitioners in the field may be well versed in reading log scale plots, few are so versed as to be able to estimate relative areas with the ease that a linear scale affords. The human visual system is tuned to home in on area differences, which is complementary when the vertical scale is linear but misleading when it is logarithmic; the reader must attempt to ignore or compensate for these cues, and the actual degree of difference between two curves on a log plot depends on its overall range. The large dynamic range representable in log plots can also hide significant features by compressing them -- a factor-of-two change covers only 6\% of the range of a 5-decade log plot, for instance.

The argument in favor of the logarithmic scale goes something like this: The sensitivity of the instruments span multiple orders of magnitude, so a feature in one part of the temperature range may be far smaller than one in another part of the temperature range and still be significant. For instance, the sensitivity of AIA at 1~MK is $\sim$20 times that at 10~MK so that, all else being equal, a feature at 10~MK needs 20 times the emission measure to have the same level of detectability. A linear scale can therefore cover `sins' of a DEM algorithm if features in the low-sensitivity temperature range dwarf those in the high-sensitivity range, making them appear tiny on the plot.

The reality is not quite as simple as that, however: errors scale as the square root of the signal, so a factor of $\sim$10 improvement in signal only leads to a factor of $\sim$3 decrease in error. Moreover, the broad and sometimes multimodal nature of the response functions means that the DEM at widely separated temperatures (with large sensitivity differences) can be coupled in unexpected ways. The algorithms are, by necessity, constructed to minimize spurious peaks with large emission measure; this both mitigates the scenario described above and it means that such primary features can be assigned some degree of trust. Secondary features, on the other hand, can often be added, removed, or shifted by making small alterations to the primary feature; determining their true significance requires an in-depth analysis, even if they occur where the instrument sensivity is high and the $\chi^2$ of the reconstruction is good.


\bibliographystyle{aasjournal}
\bibliography{simple_reg_dem_paper}

\end{document}